\documentclass{article}
\usepackage{arxiv}
\usepackage[utf8]{inputenc} 
\usepackage[T1]{fontenc}    
\usepackage{booktabs}       
\usepackage{amsfonts}       
\usepackage{nicefrac}       
\usepackage{microtype}      
\usepackage{textcomp}       
\usepackage{mathtools}
\usepackage{bm}             

\usepackage{tikz}
\usetikzlibrary{positioning}

\usepackage{times}
\usepackage{graphicx}
\usepackage{subcaption}                 
\usepackage{amssymb}
\usepackage{url}
\usepackage{hyperref}
\usepackage{geometry}
\usepackage{amsmath}
\usepackage{physics}
\usepackage{verbatim}                   
\usepackage[ruled,vlined]{algorithm2e}  
\usepackage{fancyhdr}
\usepackage{cleveref}   

\graphicspath{{figures/}}

\fancypagestyle{firststyle}
{
   \fancyhf{}
   \chead{}
}

\title{Evaluation of block encoding for sparse matrix inversion using QSVT}
\author{Leigh Lapworth \\
\\
Rolls-Royce plc \\
Derby, UK \\
\today
\\
\\
leigh.lapworth@rolls-royce.com  \\
}

\begin{document}

\maketitle
\begin{abstract}
Three block encoding methods are evaluated for solving linear
systems of equations using QSVT (Quantum Singular Value Transformation).
These are \textsc{arcsin}, \textsc{fable} and \textsc{prepare-select}. 
The performance of the encoders is evaluated using a suite of 30
test cases including 1D, 2D and 3D Laplacians and 2D CFD matrices.
A subset of cases is used to characterise 
how the degree of the polynomial approximation to $1/x$ 
influences the performance of QSVT.
The results are used to guide the evaluation of QSVT as the 
linear solver in hybrid non-linear pressure correction and
coupled implicit CFD solvers.
The performance of QSVT is shown to be resilient to 
polynomial approximation errors.
For both the pressure correction and coupled solvers,
error tolerances of $10^{-2}$ are more 
than sufficient in most cases
and in some cases $10^{-1}$ is sufficient.
The pressure correction solver also allows subnormalised condition numbers, $\kappa_s$,
as low as half of the theoretical values to be used.
This resilience reduces the number of phase factors needed and, in turn, 
reduces the time to generate the factors and emulate QSVT.
\textsc{prepare-select} encoding relies on a unitary decomposition,
e.g. Pauli strings, that has significant classical preprocessing
costs. Both \textsc{arcsin} and \textsc{fable} have much
lower costs, particularly for coupled solvers.
However, their subnormalisation factors, which are based on the rank
of the matrix, can be many times higher than \textsc{prepare-select}
leading to more phase factors being needed.
For both the pressure correction and coupled CFD calculations,
QSVT is more stable than previous HHL results due to polynomial
approximation errors only affecting long wavelength CFD errors.
Given that lowering $\kappa_s$ increases the 
success probability, optimising the performance of QSVT within a
CFD code is a function of the number QSVT phase factors, 
the number of non-linear iterations and the number of shots.
Although phase factor files can be reused, 
the time taken to generate them impedes 
scaling QSVT to larger test cases.
\end{abstract}

%
\section{Introduction}
\label{sec-intro}

With national quantum computing programmes, 
e.g. \cite{uk_quantum_missions}, increasing their focus on
error-corrected devices, it is important for end-users to
characterise and understand the likely performance of
universal quantum algorithms.
Previous work \cite{lapworth2022hybrid, lapworth2022implicit}
emulated the performance of the HHL
(Harrow-Hassidim-Lloyd) algorithm \cite{harrow2009quantum} for
two classes of Computational Fluid Dynamics (CFD) solver.
Both involve the solution of a sequence of linear systems that dominate
the run-time of classical solvers and are the most likely
candidates for quantum advantage.
With the appropriate number of eigenvalue qubits, HHL was shown to
reproduce the classical solutions. 
However, the decomposition of the CFD matrices into
Linear Combinations of Unitaries (LCU) \cite{childs2012hamiltonian}
based on tensor products of Pauli operators required classical preprocessing
with high computational costs - much higher than the classical CFD
code.
The classical preprocessing costs were most significant for the 
implicit CFD solver which transfers all matrix solutions to the
quantum computer, leaving only matrix assembly on the classical
computer.

An alternative to Trotterisation \cite{trotter1959product}
used in the HHL evaluations is qubitization
\cite{childs2012hamiltonian, babbush2018encoding, berry2015simulating, Low2019hamiltonian}.
Here, the LCU is used to encode the CFD matrix, $A$, by loading the
LCU coefficients into a \textsc{prepare} register and the unitaries
into a separate \textsc{select} register.
The benefit of qubitization is that it gives an exact encoding of
the CFD matrix up to a scale factor, thus, avoiding the 
approximation errors of Trotterization \cite{wiebe2021theory}.
Whilst qubitization can be used in Quantum Phase Estimation 
\cite{babbush2018encoding} and, hence, HHL, the additional
qubits needed for the prepare register add a significant overhead.

Quantum Singular Value Transformation (QSVT)
\cite{martyn2021grand, gilyen2019quantum, dong2021efficient} 
directly applies a polynomial function to the encoded matrix.
If a sufficiently good polynomial approximation of $1/x$ is used then
QSVT can encode $A^{-1}$ to within a user-defined accuracy and
condition number.
The asymptotic query complexity of QSVT,
$\mathcal{O}(\kappa\log (\kappa/\epsilon))$ is better than
the computational complexity of HHL, $O(log(N)s^2\kappa^2/\epsilon)$.
QSVT also has the advantage of needing a single additional qubit for 
the signal processing register.

Whilst QSVT performs well with \textsc{prepare-select} encoding, as will be shown, it
does not address the explosion in the number unitaries for implicit 
CFD matrices and the resulting classical preprocessing costs.
Direct encoding of a matrix uses a query oracle to load the
entries in the matrix 
\cite{lin2022lecture, clader2022quantum, Low2019hamiltonian, childs2017quantum,chakraborty2018power}.
These have small classical preprocessing costs but encoding 
circuit depths can scale with $\mathcal{O}(MN)$ for an 
$M \times N$ matrix.
The circuit depths can be significantly reduced for encoders
that target matrices with specific structures 
\cite{camps2022explicit, motlagh2023generalized}.
The \textsc{fable} scheme \cite{camps2022fable} provides a general algorithm
for reducing circuit depth which relies on transformation and then
cancellation of rotation gates.
Matrices that have a pattern of repeated values can
be efficiently loaded using oracles based on indexing the entries
according to how many times they have been repeated 
\cite{sunderhauf2023block}.

This work examines three matrix encoding techniques within the framework 
of QSVT as the linear equation solver for a non-linear CFD solver.
The first is an \textsc{arcsin} variant of the matrix query 
oracle that uses $\sin^{-1}$ rather than $\cos^{-1}$ based encoding. 
This only encodes the non-zero elements of the matrix.
The second is the \textsc{fable} encoding \cite{camps2022fable} and the third is 
\textsc{prepare-select} encoding.
Future work will consider encoding based on repeated values.
In addition to the CFD matrices, the encoders are investigated using a range 
of 1D, 2D and 3D Laplacian operators. 
These give matrices with a range of condition numbers that 
are used to characterise the properties of the different encoders.

The work is presented as follows.
The test cases and their condition numbers are presented first as an
area of interest is how the subnormalisation of each encoder affects the 
{\it effective} condition number, $\kappa_s$, for the QSVT phase factors.
The encoding techniques are then described and their high-level 
circuit costs are evaluated.
Next the QSVT algorithm is described, including the subnormalisation 
of the encoded matrix and the calculation of the phase factors. 
The influence of the subnormalisation on the
number of phase factors is assessed.
The influence on the precision of the phase factors on the accuracy of the
QSVT solution relative to the classical solution is evaluated for a subset
of the matrices.
Finally, results for QSVT as a linear solver within a non-linear CFD
solver are presented and conclusions drawn.
All circuit diagrams use big-endian ordering.

%
\section{Test cases}
\label{sec-testcases}

\begin{figure}[ht]
  \centering
  \captionsetup{justification=centering}
  \includegraphics[clip, trim=2.0cm 2.5cm 3.cm 3.0cm,width=0.60\textwidth]{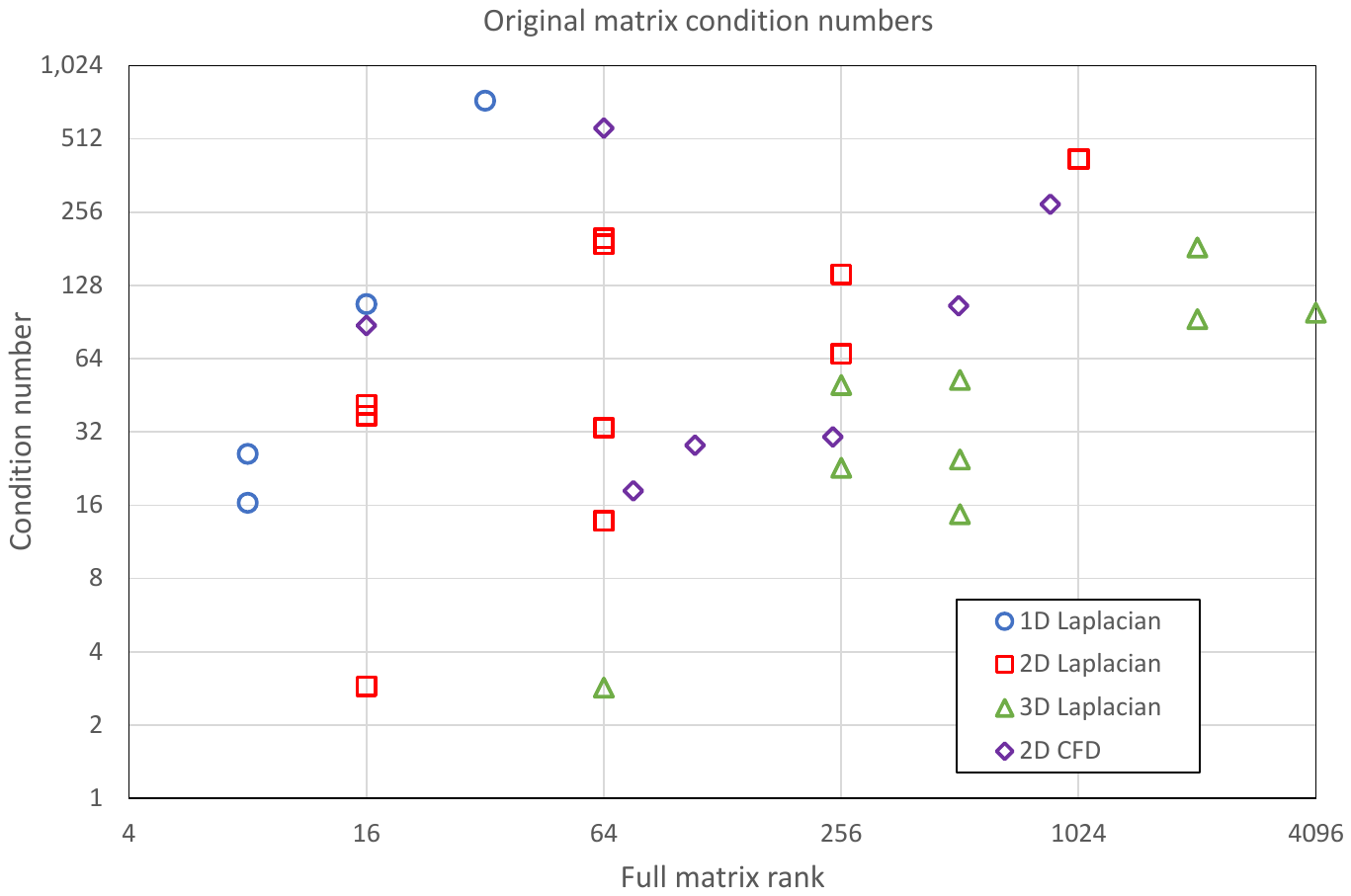}
  \caption{Condition numbers for the test matrices. Computed for original 
  non-Hermitian matrices using GSL \cite{gough2009gnu}.}
  \label{fig-kappa}
\end{figure}

The focus of this evaluation is the solution of a linear system 
within an outer non-linear solver for applications such as 
Computational Fluid Dynamics (CFD). 
Three types of test case are used:
\begin{itemize}
\item {\bf Laplacian} - 1, 2 and 3 dimensional Laplacian matrices generated
       using the L-QLES
       \footnote{\href{https://github.com/rolls-royce/qc-cfd/tree/main/L-QLES}{https://github.com/rolls-royce/qc-cfd/tree/main/L-QLES}}
       framework \cite{lapworth2024qles}.
\item {\bf Semi-Implicit CFD} -  2D pressure correction matrices
      \footnote{\href{https://github.com/rolls-royce/qc-cfd/tree/main/2D-Cavity-Matrices}{https://github.com/rolls-royce/qc-cfd/tree/main/2D-Cavity-Matrices}}
      from the SIMPLE CFD cavity solver \cite{lapworth2022hybrid}.
\item {\bf Implicit CFD} - 2D implicit matrices from the coupled CFD cavity solver
       \cite{lapworth2022implicit}.
\end{itemize}

\Cref{fig-kappa} shows the condition numbers for a range of matrices.
The values plotted are listed in \Cref{tab-kappa-all} in \Cref{app-sec-cases}.
Whilst CFD matrices can be generated for a range of meshes, there is
limited control over the resulting condition numbers.
The Laplacian cases allow a wider range of condition numbers and matrix
dimensions to be generated. This assists in the evaluation of matrix
inversion using QSVT where the condition number is the primary factor
in the number of phase factors needed.

In some of the small circuit illustrations in \Cref{sec-matenc}, the
following matrix is used. This can be generated by the L-QLES framework
but is degenerate and, hence, does not appear in \Cref{fig-kappa}.

\begin{equation}
  \frac{1}{2}
  \begin{pmatrix*}[r]
    2  & -1 & 0  & 0  & 0  & 0  & 0  & -1\\
    -1 & 2  & -1 & 0  & 0  & 0  & 0  & 0 \\
    0  & -1 & 2  & -1 & 0  & 0  & 0  & 0 \\
    0  & 0  & -1 & 2  & -1 & 0  & 0  & 0 \\
    0  & 0  & 0  & -1 & 2  & -1 & 0  & 0 \\
    0  & 0  & 0  & 0  & -1 & 2  & -1 & 0 \\
    0  & 0  & 0  & 0  & 0  & -1 & 2  & -1\\
    -1 & 0  & 0  & 0  & 0  & 0  & -1 & 2 \\
  \end{pmatrix*}
  \label{eqn-8x8-Laplace-rep}
\end{equation}

%
\section{Block encoding}
\label{sec-matenc}

Block encoding a matrix $A \in \mathbb{R}^{N \times N}$ with $N=2^n$ entails creating a 
unitary operator such that:

\begin{equation}
  U_A = 
  \begin{pmatrix}
    A/s & * \\
    *        & *
  \end{pmatrix}
  \label{eqn-UA}
\end{equation}

where $s$ is the subnormalisation constant 
\cite{clader2022quantum, lin2022lecture}.
The blocks denoted by $*$ are a result of the encoding and are
effectively {\it junk}. 
Note that the encoding requires $||A||_{max} \le 1$, any
scaling to achieve this is independent of the subnormalisation
factor.
The application of the encoding unitary
is such that:

\begin{equation}
   \frac{A}{s} = (\bra{0^m} \otimes I_n) U_A (\ket{0^m} \otimes I_n)
   \label{eqn-encode01}
\end{equation}

Direct matrix encoders rely on a \textsc{query-oracle}, $O_A$ that returns the 
value for each entry $(i,j)$ in the matrix \cite{lin2022lecture}:

\begin{equation}
   O_A \ket{0}\ket{i}\ket{j} =
   \left(
   a_{ij}\ket{0} + \sqrt{1-|a_{ij}|^2}\ket{1}
   \right)
   \ket{i}\ket{j}
   \label{eqn-OA-cos}
\end{equation}

\begin{figure}[ht]
  \centering
  \captionsetup{justification=centering}
  \includegraphics[width=0.50\textwidth]{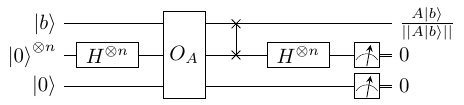}
  \caption{Schematic of block encoding circuit for $A/s$ from
           \cite{camps2022fable}.}
  \label{fig-UA-circ}
\end{figure}

A circuit implementation of the block encoding unitary is shown
in \Cref{fig-UA-circ} from \cite{camps2022fable}.

\begin{figure}[ht]
  \centering
  \captionsetup{justification=centering}
  \includegraphics[width=0.90\textwidth]{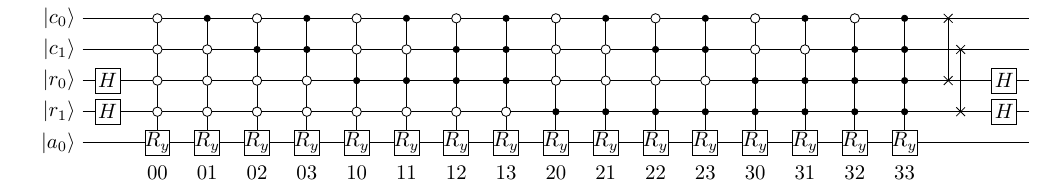}
  \caption{Block encoding circuit for a 4x4 matrix. Labels below the 
           circuit indicate the row and column indices for the 
           controlled rotations.}
  \label{fig-fable_4x4_orig}
\end{figure}

The \textsc{query-oracle} consists of \textit{row} and \textit{column} registers,
each with $n = \log_2 N$ qubits. 
Multi-controlled rotations are applied to an ancilla qubit.
The multiplexing of controls are such that the correct entry
$a_{ij}$ is loaded into the encoding block and 
$\sqrt{1-|a_{ij}|^2}$ is loaded into the junk block.
\Cref{app-2x2enc} gives an explicit derivation of the block encoding
of a 2x2 matrix.
\Cref{fig-fable_4x4_orig} shows the full encoding circuit 
for a 4x4 matrix.

For this encoding circuit, the subnormalisation factor
is $s = 2^n$ and the probability of measuring all the ancilla qubits in the $\ket{0}$ state is \cite{lin2022lecture}:

\begin{equation}
    P(0^{n+1}) = \frac{1}{s^2} ||A\ket{b}||^2 
           = \frac{1}{s^2} \bra{b}A^{\dagger} A \ket{b}
\end{equation}

where $\ket{b}$ is the right hand side (RHS) state of the linear system
$A \ket{x} = \ket{b}$ to be solved.

%
\subsection{\textsc{arcsin} encoding}
\label{subsec-arcsin}
In order for the circuit in \Cref{fig-fable_4x4_orig} to correctly
encode $A$, the rotation angles, $\theta$ must be set so that:
$\theta_{ij} = 2 \cos^{-1} (a_{ij})$.
For sparse matrices, this leads to a large circuit overhead as
all the zero entries have rotation angles of $\pi$.
Sparse matrix encoders can be thought of
as equivalent to the Compressed Sparse Row (CSR) storage format.
If $S_r$ is the maximum number of entries per row, the column oracle
must convert the CSR index $0 \le s \le S_r$ for each row into the 
corresponding column index. 
This generally requires an additional work register \cite{lin2022lecture}.

An alternative approach is presented in \Cref{app-2x2enc}
where the rotation angles are calculated by
$\theta_{ij} = 2 \sin^{-1} (a_{ij})$.
This requires only the addition of a Pauli X gate
on the ancilla rotation qubit at the end of the encoding circuit.
This is equivalent to projecting the ancilla qubit into
the $\ket{1}$ computational basis and the encoding oracle,
with angles based on \textsc{arcsin}, becomes:

\begin{equation}
   O_A \ket{0}\ket{i}\ket{j} =
   \left(
   \sqrt{1-|a_{ij}|^2}\ket{0} + a_{ij}\ket{1} 
   \right)
   \ket{i}\ket{j}
   \label{eqn-OA-sin}
\end{equation}

There are now only as many multi-controlled rotations 
as there are non-zeros in the sparse matrix.
However, the column indexing requires the same number of qubits
in the column register.

%
\subsubsection{Circuit trimming}
\label{subsubsec-arcsin-trim}
\textsc{arcsin} encoding enables two options for reducing the depth of the encoding circuit.
The first is to set rotation angles below a small threshold to zero, hence removing the
corresponding operation from the ciruit. This is the same as setting small entries in the
matrix to zero.

\begin{figure}[ht]
  \centering
  \captionsetup{justification=centering}
  \includegraphics[width=0.25\textwidth]{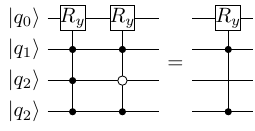}
  \caption{Coalescing multi-qubit controlled gates with a Hamming distance of 1 where
   the $R_y$ gates have the same rotation angle.}
  \label{fig-mcon-gate-collapse}
\end{figure}

The second approach is to coalesce two multi-controlled operations which have the same 
rotation angle and where the two  multiplexing strings are identical except for 
one qubit where they are bit-flipped.
As shown in \Cref{fig-mcon-gate-collapse} this is equivalent to the bit patterns for
the multiplexing strings having a Hamming distance of 1.
Since the multi-controlled rotations commute, the circuit can be reordered to find pairs
that can be coalesced. The pairing can also applied recursively to pair previously
coalesced operations. 

\begin{figure}[ht]
  \centering
  \captionsetup{justification=centering}
  \includegraphics[width=0.8\textwidth]{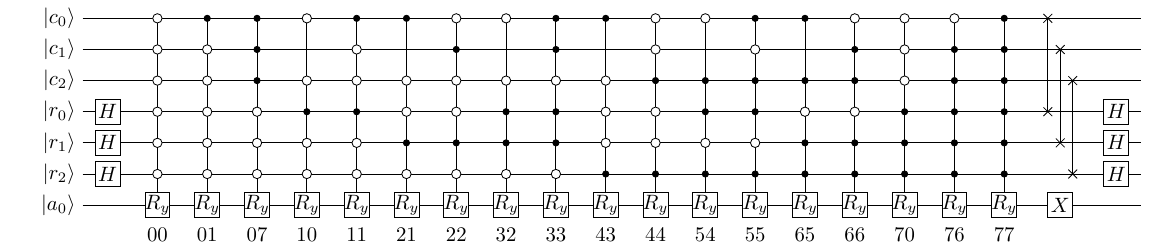}
  \caption{\textsc{arcsin} encoding circuit for 8x8 tri-diagonal matrix with coalesence of
           equal valued off-diagonal rotations on rows 1 to 6.}
  \label{fig-fable_8x8_1D_Laplace_orig_sin_trim}
\end{figure}

\Cref{fig-fable_8x8_1D_Laplace_orig_sin_trim} shows a trimmed encoding
circuit for the matrix in \Cref{eqn-8x8-Laplace-rep}.
For example, on the second row of the matrix, the entries 1,0 are 1,2
are both equal and 1 bit flip apart. These are coalesced into the
entry at 1,0. 
The subsequent rows, except the last, can be similarly trimmed.
The current implementation always retains the operation with the lowest
bit value. 
For this matrix, retaining entry 1,2 on row 2 and entry 3,2 on row 4
would allow further trimming.

More generally, approximations can be used to set values that are sufficiently 
close to have the same value.
Caution is needed when approximating small values by zero.
The value may be small because it makes a negligible contribution, or it may be zero
because it represents a small scale in a multiscale discretisation.

An advantage of the \textsc{arcsin} encoding is that trimming operations involve a direct correlation
between the matrix entries and the rotation angles. Whilst setting matrix entries to zero or 
equal to others may assist \textsc{fable} and \textsc{prepare-select} encoding, 
their routes to circuit trimming involved derived quantities.
More importantly, the changes to the matrix are known {\it a priori} and some of the 
effects may be able to be mitigated.
For example, matrices resulting from a finite volume discretisation have a conservation 
property where the sum of the entries on a row is zero.
Where matrix entries have been modified to reduce the circuit depth, additional changes can
be made to recover the conservation property.

%
\subsection{\textsc{fable} encoding}
\label{subsec-fable}
The \textsc{fable} method \cite{camps2022fable} is derived from the matrix \textsc{query-oracle} in
\Cref{eqn-OA-cos} and illusrated in \Cref{fig-fable_4x4_orig}.
The key insight in \textsc{fable} is to re-express the multi-controlled rotations as an
interleaving sequence of uncontrolled rotations and CNOT gates in Gray code ordering.
Further, if $\bm{\theta}$ is the $N$ dimensional vector of rotation angles used in $Q_A$, 
and $\bm{\hat{\theta}}$ the angles for the uncontrolled rotations in Gray code order,
Then the angles are related by:

\begin{equation}
  \bm{\theta} = 2^n H^{\otimes 2n} P_G \bm{\hat{\theta}}
  \label{eqn-fable-angles1}
\end{equation}

Where $H$ is the Hadamard gate and $P_G$ is the permutation matrix that transforms
binary ordering to Gray code ordering.
\Cref{eqn-fable-angles1} can be rewritten as:

\begin{equation}
  \bm{\hat{\theta}} = \frac{1}{2^n} P_{G}^{\dagger} H^{\otimes 2n} \bm{\theta}
  \label{eqn-fable-angles2}
\end{equation}

Since the inverse mapping $P_{G}^{\dagger}$ can be easily computed, the angles $\bm{\hat{\theta}}$ 
can be directly evaluated.

The premise of \textsc{fable} is that the solution of \Cref{eqn-fable-angles2} leads to a large 
number of angles that are zero or close to zero.
Removing all angles with $\hat{\theta_j} \le \delta_c$ leads to a block encoding error
\cite{camps2022fable}:

\begin{equation}
  ||A - s\hat{A} ||_2 \le N^3 \delta_c + \mathcal{O} (\delta_{c}^{3})
  \label{eqn-fable-angles3}
\end{equation}

where $\hat{A}$ is the encoding from \Cref{eqn-encode01} with $U_A$ based on the
thresholding of small angles.

\begin{figure}[ht]
  \centering
  \captionsetup{justification=centering}
  \includegraphics[width=.9\textwidth]{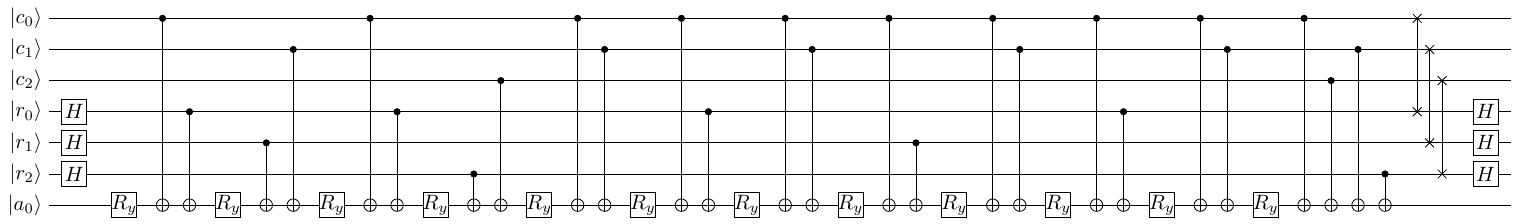}
  \caption{\textsc{fable} encoding circuit for 8x8 tri-diagonal matrix.}
  \label{fig-fable_8x8_1D_Laplace_trim2}
\end{figure}

Each rotation gate that is removed brings two CNOT gates together. 
Where there are sequences of zero-valued rotations, the resulting string of CNOTS may
enable cancellation of pairs that operate on the same qubits.
See Section B of \cite{camps2022fable}. 
\Cref{fig-fable_8x8_1D_Laplace_trim2} shows the \textsc{fable} circuit for the same
8x8 tridiagonal matrix as shown in \Cref{fig-fable_8x8_1D_Laplace_orig_sin_trim}.

Note that the 8x8 matrix used in the figures has entries of 1 along the diagonal 
and -0.5 along the off diagonals, resulting in 52 of the 64 values of $\bm{\hat{\theta}}$
being zero, without the need for thresholding small values.
When the number of zeros in $\bm{\hat{\theta}}$  is small, 
the conversion from multi-controlled
rotations to uncontrolled rotations and CNOT gates may have benefits when the circuit is
transpiled to native gates.

%
\subsection{\textsc{prepare-select} encoding}
\label{subsec-prep-sel}

\textsc{prepare-select} encoding follows from expressing the matrix to be encoded as a 
Linear Combination of Unitaries (LCU)
\cite{childs2012hamiltonian, berry2015simulating,kothari2014efficient, 
berry2018improved, babbush2018encoding, Low2019hamiltonian}:

\begin{equation}
  A_H = \sum_{i=0}^{M-1} \alpha_i U_i
  \label{eqn-lcu01}
\end{equation}

Where $M$ is the number of entries in the LCU and $A_H$ is bipartite Hermitian matrix
with $A$ in the top right block and $A^{\dagger}$ in the bottom left block.
If $A$ is already Hermitian, $A_H = A$.
Typically, each unitary is a product of Pauli matrices.
\textsc{prepare-select} encoding requires $\alpha_i > 0$ which can be achieved by taking the
signs of any negative coefficients into the corresponding unitary $U_i$.
The resulting encoding unitary is:

\begin{equation}
  U_A = (P^{\dagger} \otimes I^n) S (P \otimes I^n) 
  \label{eqn-prep-sel01}
\end{equation}

The prepare operator, $P$, is defined by its action of the $\ket{0}$ state:

\begin{equation}
  P\ket{0} = \sum_{i=0}^{M-1} \sqrt{\frac{\alpha_i}{s}} \ket{i}
  \label{eqn-prep01}
\end{equation}

where $s = \norm{\alpha}_1 = \sum_{i=0}^{M-1} |\alpha_i|$ is the $L_1$ norm of the coefficients
and is the subnormalisation constant.
Essentially, $P$ is a state loader for the coefficients $\sqrt{\frac{\alpha_i}{s}}$.
In this work, the a binary tree data loader is used 
\cite{mottonen2004transformation, araujo2021divide}.
In gereral, the number of unitaries, $M$, in the LCU is not a power of 2 and the 
vector of coefficients passed to the state loader must be padded with zeros.

The select operator, $S$ is:

\begin{equation}
  S = \sum_{i=0}^{M-1} \ket{i}\bra{i} \otimes U_i
  \label{eqn-sel01}
\end{equation}

\begin{figure}[ht]
  \centering
  \captionsetup{justification=centering}
  \includegraphics[width=0.75\textwidth]{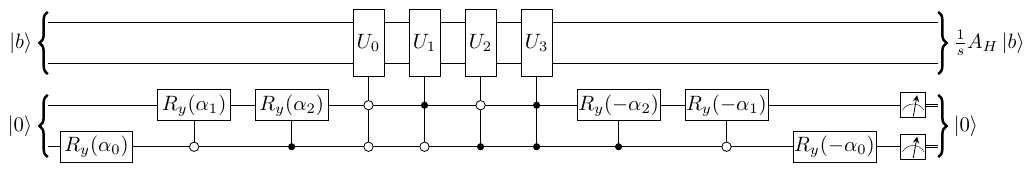}
  \caption{Circuit for encoding a 4x4 matrix using \textsc{prepare-select}.}
  \label{fig-prep-sel-4x4}
\end{figure}

\Cref{fig-prep-sel-4x4} shows the \textsc{prepare-select} circuit for encoding 
a 4x4 matrix with 4 entries in its LCU.

%
\subsection{Comparison of encoding techniques}
\label{subsec-enc-compare}
Performing a like-with-like comparison of the three encoding schemes
is not straightforward as they result in different styles of circuits.
Hence, only an order of magnitude analysis is undertaken.
For both \textsc{arcsin} and \textsc{fable} encoding, the number of rotation gates is used.
This ignores the fact that \textsc{arcsin} uses multi-controlled rotations and
\textsc{fable} uses single qubit rotations. The number of CNOT gates in \textsc{fable} circuit
is also ignored.
For \textsc{prepare-select}, the number of terms in the LCU is used. This
is multiplied by a factor of 3 as the Prepare circuit, its adjoint
and the Select circuit have the same number of operations.
Hadamard and swap gates that scale with the number of qubits are
not counted.

\begin{figure}[ht]
  \centering
  \captionsetup{justification=centering}
  \includegraphics[clip, trim=2.0cm 2.5cm 3.cm 3.0cm,width=0.60\textwidth]{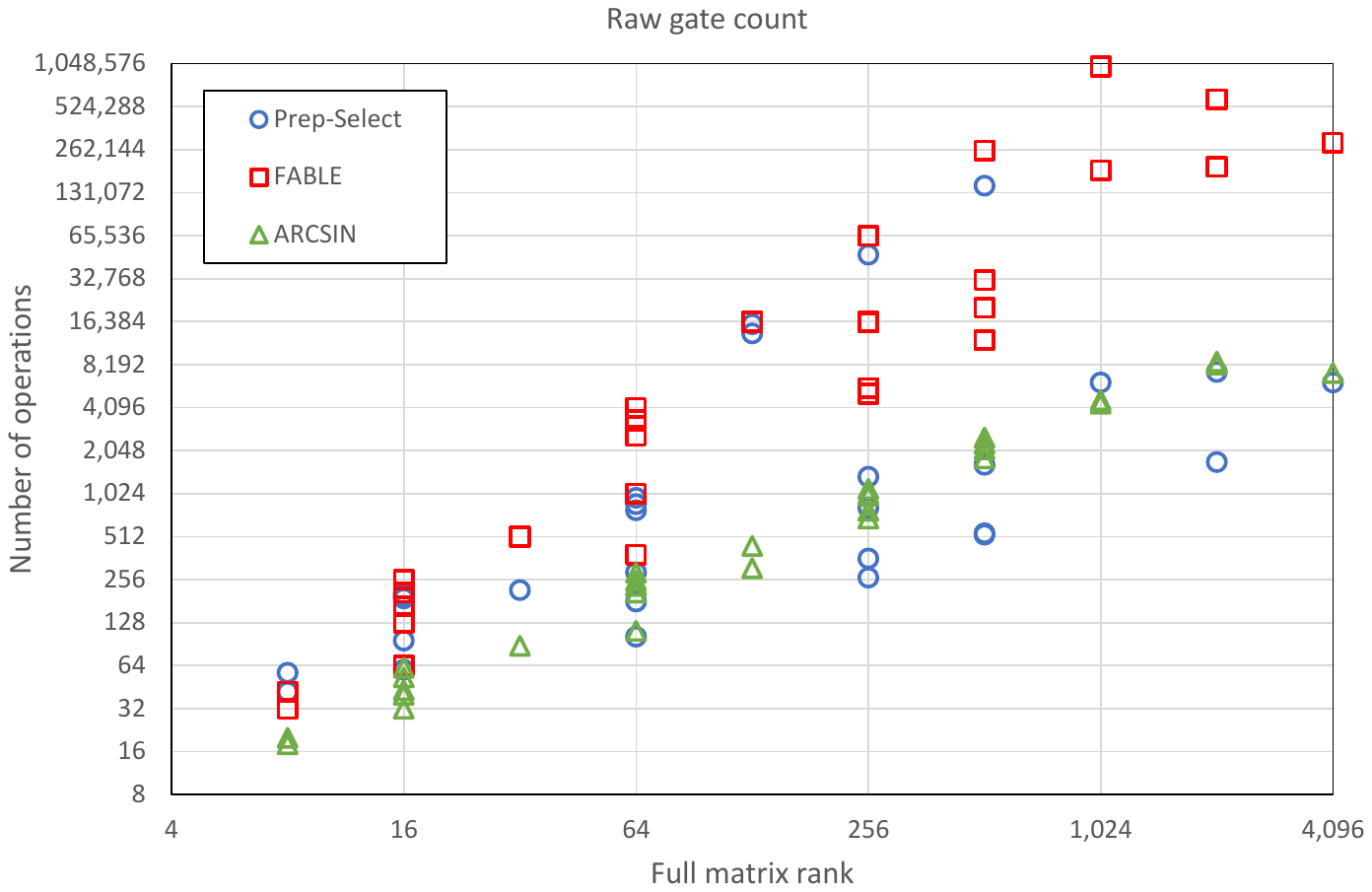}
  \caption{Number of encoding operations. Rotation gates for \textsc{fable} and \textsc{arcsin}
           encoding and number of unitaries in the LCU for \textsc{prepare-select}.}
  \label{fig-nrot-raw}
\end{figure}

\Cref{fig-nrot-raw} shows the operation counts for each of the cases
in \Cref{fig-kappa}. The plotted values are listed in \Cref{tab-nopers-all}
in \Cref{app-subsec-nopers}.
By construction, the number of operations in the \textsc{arcsin} encoding
scales with the number of non-zeros in the matrix.
Most of the \textsc{prepare-select} cases have similar scaling except for
those for the coupled CFD matrices which have close to 
$\mathcal{O}(N^2)$ scaling. The \textsc{fable} results show the largest
scatter.

\begin{figure}[ht]
  \centering
  \captionsetup{justification=centering}
  \includegraphics[clip, trim=2.0cm 2.5cm 3.cm 3.0cm,width=0.60\textwidth]{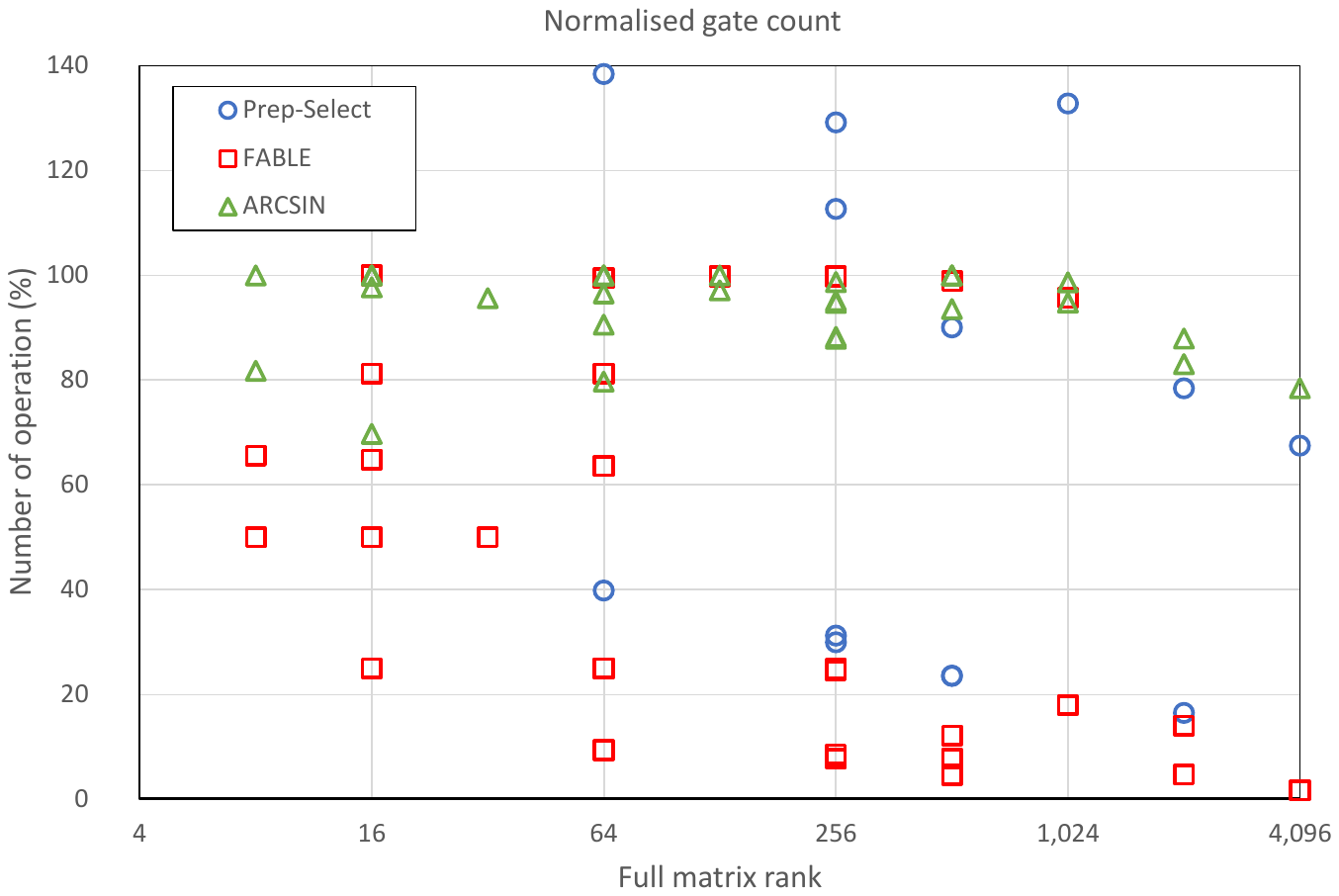}
  \caption{Normalised operation counts for the data in \Cref{fig-nrot-raw}.}
  \label{fig-nrot-norm}
\end{figure}

An alternative view of the number of operations is shown in
\Cref{fig-nrot-norm}.
Here, the \textsc{prepare-select} and \textsc{arcsin} encodings are normalised by
the number of non-zeros and the \textsc{fable} encoding by $N^2$. 
For the Laplacian cases, \textsc{fable} encoding is very effective on
the larger cases - although the raw number of operations remains
very high. The \textsc{arcsin} circuit trimming reduces the operation 
count for many but not all of the Laplacian cases.
However, it is not as effective as \textsc{fable} at doing so.
As with \textsc{fable}, there are Laplacian cases where \textsc{prepare-select}
encoding uses only a small percentage of Pauli strings.
There are also cases where the number of \textsc{prepare-select} operations
far exceeds the number of non-zeros.
Several of these are not shown in \Cref{fig-nrot-norm}
including all of the coupled CFD matrices.
For the CFD matrices, the number of \textsc{fable} and \textsc{arcsin}
operations are both close to 100\%. 
By construction, they cannot be larger. As shown in
\Cref{fig-nrot-norm} this does not mean they have the same
number of operations.

Applying the \textsc{fable} mapping, \Cref{eqn-fable-angles2}, to the
\textsc{arcsin} encoding is possible if all the zero angle rotations
are included to enable the Gray coding step to be completed.
This was tested and found to produce very similar operation
counts to the \textsc{fable} scheme.
Whether the initial vector of angles, $\bm{\theta}$,
contained a large number of zeros or a large number
angles equal to $\pi$, had only a small effect on the
number of non-zero angles in the output vector $\bm{\hat{\theta}}$.

Demultiplexing the multiplexed rotations 
\cite{shende2004synthesis,shende2005synthesis} in the
\textsc{arcsin} and \textsc{prepare-select} encoding increases
the number of rotation gates by a factor of $\mathcal{O}(\log_{2}N)$.
This has not been done as multiplexed operators are efficient to
implement in emulation.

Since, without any approximations,
\textsc{arcsin} and \textsc{fable} encoding result in the same unitary and
have the same subnormalisation factor the following analysis 
will omit \textsc{fable}, as \textsc{arcsin} encoding is more efficient to emulate.
The results labelled \textsc{arcsin} can also be read as being \textsc{fable}
results.

%
\section{Quantum Singular Value Transformation}
\label{sec-qsvt}

Quantum Singuar Value Transformation (QSVT) derives from Quantum Signal Processing 
\cite{martyn2021grand, gilyen2019phd, gilyen2019quantum, dong2021efficient} 
and applies a polynomial function $P$ to the encoded block in \Cref{eqn-UA}:

\begin{equation}
  \begin{pmatrix}
    A/s & * \\
    *        & *
  \end{pmatrix}
  \xmapsto[]{QSVT}
  \begin{pmatrix}
    P(A/s) & * \\
    *           & *
  \end{pmatrix}
  \label{eqn-qsvt01}
\end{equation}

\begin{figure}[htb]
  \centering
  \captionsetup{justification=centering}
  \includegraphics[width=0.75\textwidth]{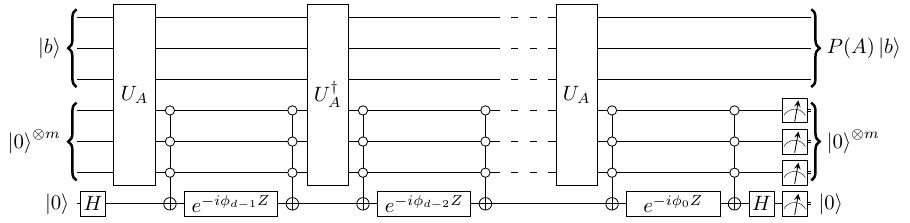}
  \caption{QSVT circuit for the transformation in \Cref{eqn-martyn32} for 8x8 matrix and
           a degree-d odd polynomial.}
  \label{fig-martyn_eval_BE}
\end{figure}

The Polynomial is implemented by interleaving applications of $U_A$ and $U_{A}^{\dagger}$
with projector controlled phase shifts as shown in \Cref{fig-martyn_eval_BE}.
For an {\bf odd} polynomial of degree $d$, the transformation is:
\begin{equation}
  S_{\vec{\phi}} =
  \Pi_{\phi_1} U \left[ \prod\limits_{k=1}^{(d-1)/2} \Pi_{\phi_{2k}} U_{A}^{\dagger} \Pi_{\phi_{2k+1}} U_{A}\right]  
  \label{eqn-martyn32}
\end{equation}

where $\Pi_{\phi}$ is a projector controlled phase shift operator:
$\Pi_{\phi} = e^{i 2 \phi \Pi}$.
The implementation of the desired polynomial transformation depends of
the the phase factors, $\phi_k$ for $k={0, \ldots, d-1}$.

%
\subsection{Subnormalisation}
\label{subsubsec-subnorm}

The subnormalisation constant, $s$ depends on the method of block encoding.
For \textsc{prepare-select},  
$s = \norm{\alpha}_1 = \sum_{i=0}^{M-1} |\alpha_i|$ where $\alpha$
is the vector of coefficients in the LCU decomposition.
For the query oracles, \textsc{fable} and \textsc{arcsin} encoding, $s=N$.

In addition, for \textsc{query-oracles}s, it must also be ensured that 
$||A||_{max} \le 1$. For solving linear systems, this is applied as an
external scaling prior to QSVT:

\begin{equation}
    \frac{A}{||A||_{max}} \ket{x} = \frac{1}{||A||_{max}} \ket{b}
    \label{eqn-rescale}
\end{equation}

Scaling both $A$ and $\ket{b}$ ensures that the solution to the original
system is recovered.

When computing the phase factors for the matrix inversion it is the condition
number $\kappa_s$ of $A/s$ that should be used. Due to the subnormalisation,
if the encoded matrix is square, this gives:
\begin{equation}
  \kappa_s = \frac{1}{|\lambda^{min}_{s}|}
  \label{eqn-qsvt-subnorm03}
\end{equation}

where $\lambda^{min}_{s}$ is the smallest eigenvalue of $A/s$.
If $A/s$ is not square, then $\lambda^{min}_{s}$ is the smallest singular value.
Note that while $\lambda^{min}_{s} = \lambda^{min}/s$ this does not,
in general, mean that $\kappa_s = s\kappa$. However, it does give:
\begin{equation}
  \frac{\kappa_s}{s} = \frac{1}{|\lambda^{min}|}
  \label{eqn-qsvt-subnorm04}
\end{equation}

For \textsc{arcsin} encoding, 
if $||A||_{max} < 1$, it is beneficial to apply the 
scaling in \Cref{eqn-rescale} as this increases $\lambda^{min}_{s}$ and
reduces $\kappa_s$ since the subnormalisation constant is unchanged.
For \textsc{prepare-select} encoding this is not needed as the scaling increases both
the minimum eigenvalue and the LCU coefficients and does not affect
the subnormalisation constant.

\begin{figure}[htb]
  \centering
  \captionsetup{justification=centering}
  \includegraphics[clip, trim=2.0cm 2.5cm 3.cm 2.5cm,width=0.60\textwidth]{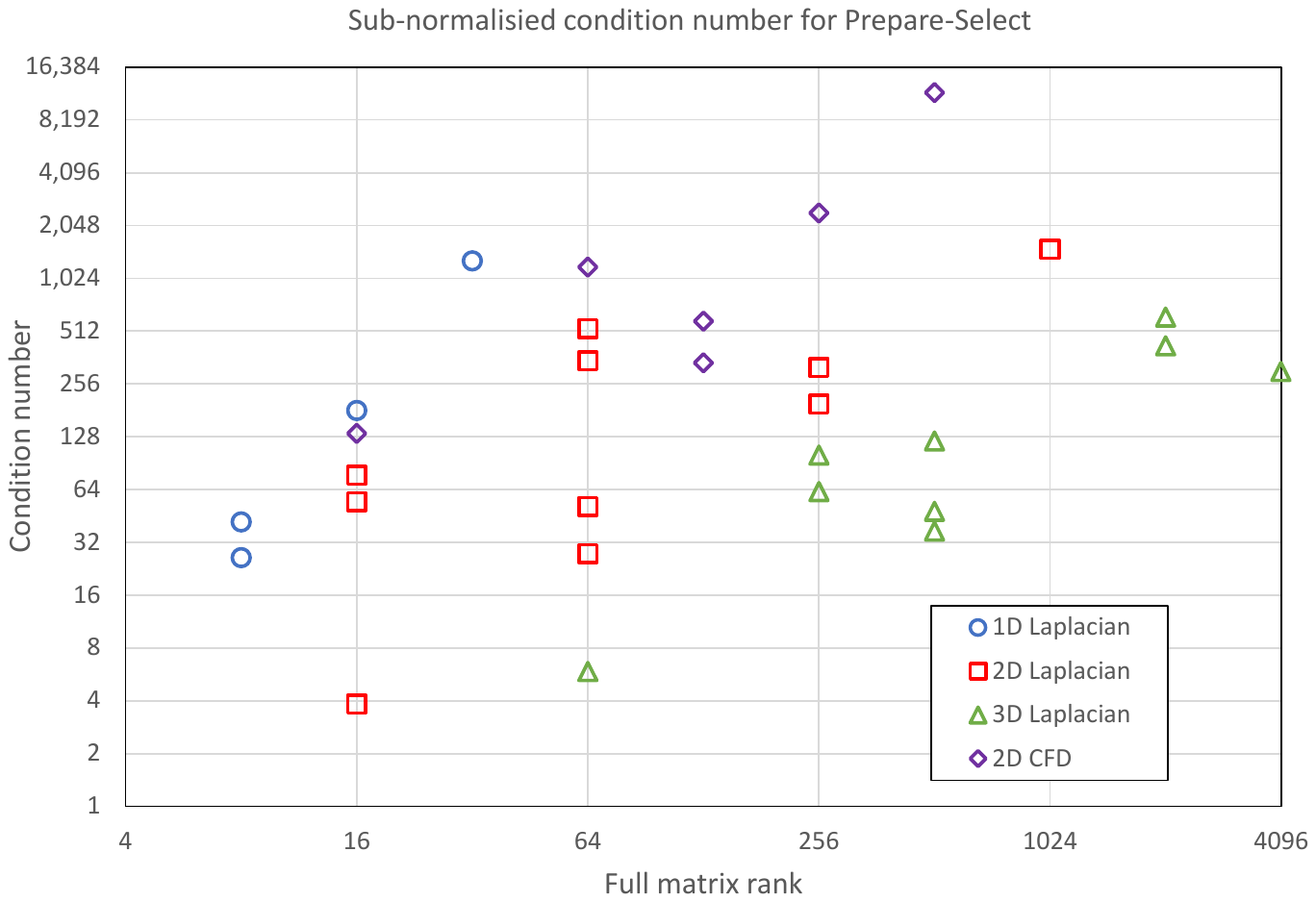}
  \caption{Condition numbers for subnormalised matrix using
  \textsc{prepare-select} encoding.}
  \label{fig-kappa-ps}
\end{figure}

\begin{figure}[htb]
  \centering
  \captionsetup{justification=centering}
  \includegraphics[clip, trim=2.0cm 2.5cm 3.cm 2.5cm,width=0.60\textwidth]{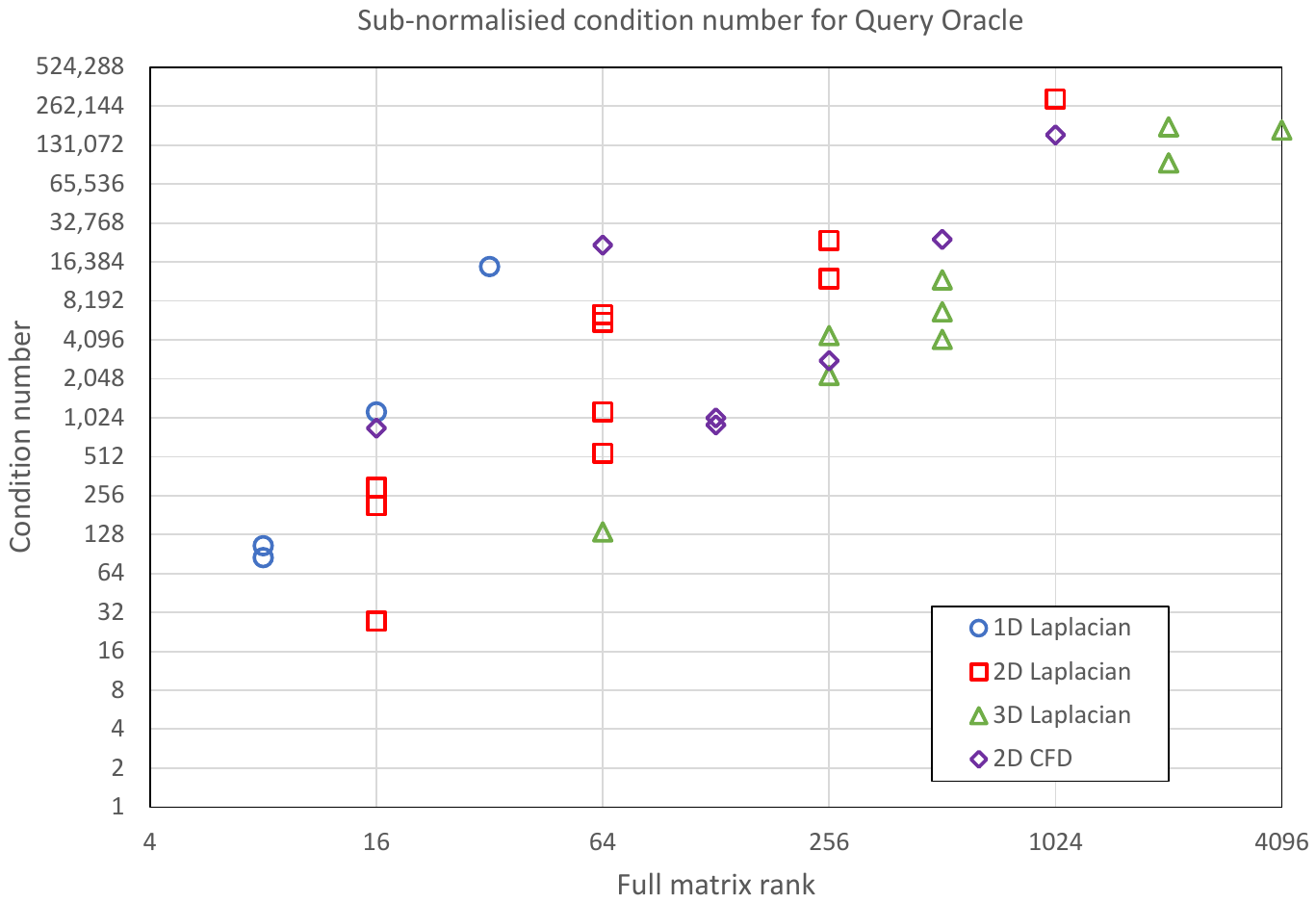}
  \caption{Condition numbers for subnormalised matrix using
  \textsc{arcsin} encoding.}
  \label{fig-kappa-qo}
\end{figure}

\Cref{fig-kappa-ps} shows the subnormalised condition numbers
for all the test cases using \textsc{prepare-select} encoding.
For the Laplacian and pressure correction CFD matrices the
scaling relative to the original matrices is $\mathcal{O}(1)$.
For the coupled CFD matrices the scaling is $\mathcal{O}(10)$ 
due to far greater number of unitaries in the LCU.

\Cref{fig-kappa-qo} shows the subnormalised condition numbers
using \textsc{arcsin} encoding. Here, the scaling is equal to the
rank of the matrix. Note that this is not a direct scaling of the
original condition number as $\kappa_s$ depends on 
$|\lambda^{min}_{s}|$.
Note also that the ranks for coupled CFD matrices are for
the encoded block for which the rank is a power of 2.
The subnormalised condition numbers for both encoding methods
are listed in \Cref{tab-kappa-all} in \Cref{app-subsec-kappa}.

%
\subsection{Scaling the QSVT solution}
\label{subsec-scale-x}

As with all QLES methods, the normalised solution state $\ket{\hat{x}}$ must
be scaled to return the correct solution to the original matrix equation.
As the matrix being scaled by subnormalisation factor, the polynomial
being applied is also scaled. 
For the Dong et al. \cite{dong2021efficient} Remez approximation used 
here, the polynomial being approximated is:
\begin{equation}
  P(x) = \frac{1}{4\kappa_s x}
  \label{eqn-qsvt-subnorm05a}
\end{equation}

Hence, if QSVT has accurately implemented $P$, the state at the end of
QSVT is:


\begin{equation}
  \ket{0^{m+1}}\otimes \ket{\hat{b}} \xmapsto[]{QSVT} 
  \ket{0^{m+1}}\otimes \frac{s}{4\kappa_s} A^{-1} \ket{\hat{b}}
  + \ket{\Phi^{\perp}}
  \label{eqn-qsvt-subnorm07}
\end{equation}

Where $\ket{\Phi^{\perp}}$ is the orthogonal \textit{junk} state.
The expectation, $E$, of measuring all the $m+1$ flag qubits in the
$\ket{0}$ state is:

\begin{equation}
  E = \left(\frac{s}{4\kappa_s}\right)^2 ||A^{-1}||_{2}^{2}
    \label{eqn-qsvt-expect}
\end{equation}

And the post measurement state in the encoding qubits is:

\begin{equation}
  \frac{1}{\sqrt{E}}\frac{s}{4\kappa_s} A^{-1} \ket{\hat{b}}
  =
  \frac{1}{\sqrt{E}}\frac{s}{4\kappa_s} \ket{\hat{x}}
  \label{eqn-qsvt-subnorm08}
\end{equation}

If the  normalised input state is $\ket{\hat{b}} = \ket{b}/||\bra{b}||_2$, 
the re-dimensionalised output state is:

\begin{equation}
  \ket{x} = \frac{4 \kappa_s}{s} \sqrt{E} ||\ket{b}||_2 \ket{\hat{x}}
\end{equation}

Whilst the expectation \Cref{eqn-qsvt-expect} has no explicit
dependence on $\epsilon$, the analysis has not accounted for the
degree to which $P$ approximates $A^{-1}$.
This influence will be analysed \Cref{subsec-success}.

%
\subsection{Computing the phase factors}
\label{subsec-qsppack}
The phase factors are calculated via the Remez method  
\cite{dong2021efficient,dong2023robust}
using the open-sourced \textsc{qsppack}
\footnote{\href{https://github.com/qsppack/QSPPACK}{https://github.com/qsppack/QSPPACK}}
software package.
Whilst the degree of the polynomial scales with 
$\mathcal{O}(\kappa \log(1/\epsilon))$,
setting the value of the degree can involve some trial and error.
Within the context of QLES, the required accuracy of solution state
also has a bearing on the degree of the polynomial.
To investigate this, a tridiagonal Toeplitz matrix is used:

\begin{equation}
  T_n(a,b,c)
  = 
  \begin{pmatrix}
    a       & c      &        &        &        &        &   \\
    b       & a      & c      &        &        &        &   \\
            & b      & a      & c      &        &        &   \\
            &        & \ddots & \ddots & \ddots &        &   \\
            &        &        &        & b      & a      & c \\
            &        &        &        &        & b      & a \\
  \end{pmatrix}
\label{eqn_toeplitz_tri01}
\end{equation}

For which, the eigenvalues are:

\begin{equation}
  \lambda_k = a - 2 \sqrt{bc} \cos(\frac{k\pi}{n+1}), k=1,\ldots,n
  \label{eqn_toeplitz_evals01}
\end{equation}

\Cref{eqn_toeplitz_evals01} enables a matrix with a specified
condition number to be created. 
To make this easier, the settings $a=1$ and $b=c$ are used.
Using \textsc{prepare-select} encoding, a 32x32 Toeplitz matrix with $\kappa=28.8$
gave a subnormalised matrix with $\kappa_s=49.98$.
Using the error relative to the classical solution, \textsc{qsppack} was used to
generate phase factors for $\kappa_s=50$ and $L_{\infty}$ errors of 
$10^{-1}$, $10^{-2}$ and $10^{-3}$. 
The degrees were adjusted to give approximately the same errors in the
$L_2$ norms of QSVT solution of the Toeplitz matrix.
The resulting degrees were 109, 229 and 359.
For this analysis, the amplitudes of right hand side vector, $\ket{b}$ 
were set from:
$b_i = 16x_{i}^{3} - 24x_{i}^{2} + 9x_{i}$ for $i=0,N-1$ 
and $x_i = 1/(N-1)$.

\begin{figure}[htb]
  \centering
  \captionsetup{justification=centering}
  \includegraphics[clip, trim=0.0cm 0.0cm 0.0cm 0.0cm,width=0.60\textwidth]{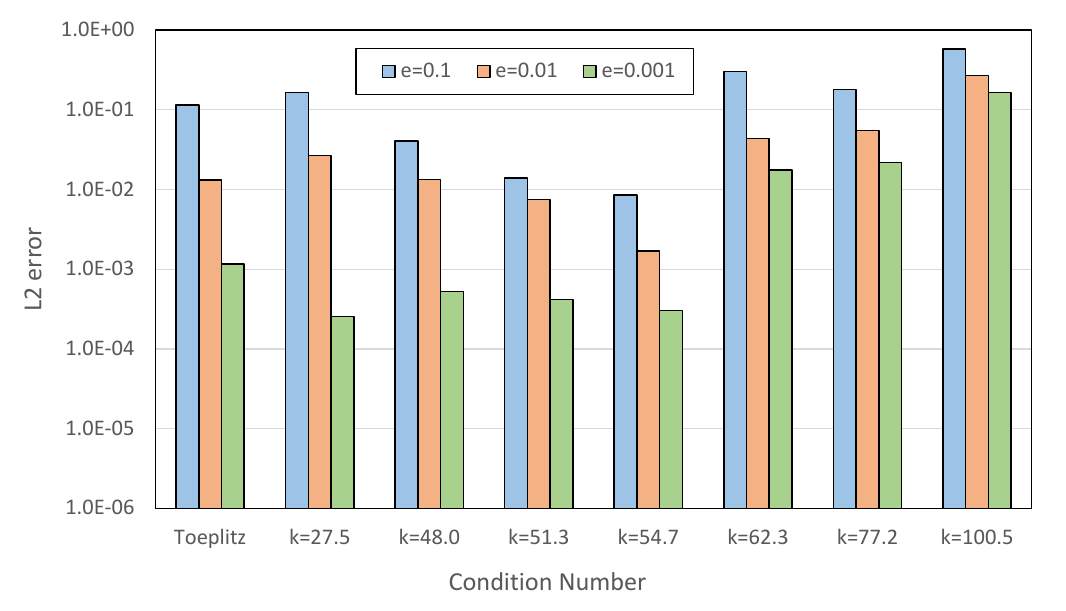}
  \caption{$L_2$ errors for range of Laplacians and \textsc{qsppack} $L_{\infty}$ error tolerances. All with 
  $\kappa_s=50$ phase factors and \textsc{prepare-select} encoding.}
  \label{fig-k50-e01-e03-l2}
\end{figure}

\begin{table}[htb]
  \centering
  \captionsetup{justification=centering}
  \begin{tabular}{l l l }
    \toprule
    $\kappa_s$ & Case name & Case index \\
    \midrule
     27.5 &	l2d\_8x8\_dddd & 8 \\
     48.0 &  l3d\_8x8x8\_ddrrdd & 19 \\
     51.3 &	l2d\_8x8\_ddrr & 9 \\
     54.7 &	l2d\_4x4\_rrrr & 7 \\
     62.3 &	l3d\_4x8x8\_dndddd & 16 \\
     77.2 &	l2d\_4x4\_nnnn & 6 \\
    100.5 &	 l3d\_4x8x8\_dnrrdd & 17 \\
    \bottomrule
    \\
  \end{tabular}
  \caption{Test cases order by subnormalised condition number
  using \textsc{prepare-select} encoding.
  Case index refers to position in \Cref{tab-kappa-all}.}
  \label{tab-k=5-cases}
\end{table}

\Cref{fig-k50-e01-e03-l2} shows the influence of \textsc{qsppack} $L_{\infty}$ tolerance
on the Toeplitz matrix for the Laplacian cases listed in 
\Cref{tab-k=5-cases}.
All the solutions are calculated using the same expression for
$\ket{b}$ as used in the Toeplitz calculations, 
and all were solved using \textsc{prepare-select} encoding.
There is some variation in the $L_2$ error levels between the cases.
As expected, the $L_2$ solver errors increase for the cases with higher condition
numbers and the benefits of more accurate phase factors becomes more marginal.
Further analysis using the RHS states generated by the L-QLES framework showed that
the $L_2$ error levels are also influenced by the eigen spectrum of the input state
which is dependent on the matrix.

\begin{figure}[htb]
  \centering
  \captionsetup{justification=centering}
  \includegraphics[clip, trim=0.0cm 0.0cm 0.0cm 0.0cm,width=0.60\textwidth]{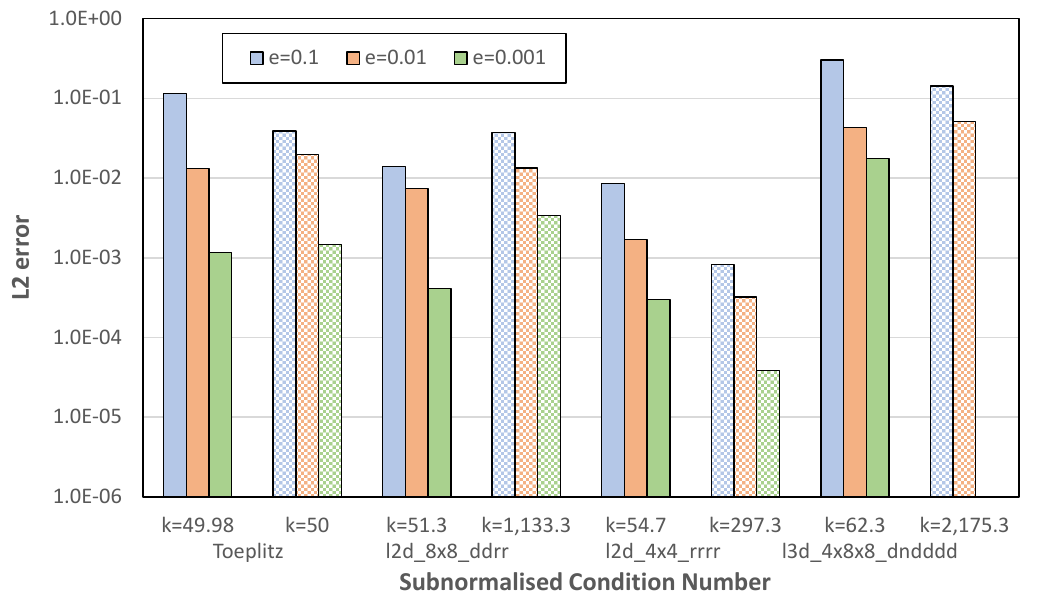}
  \caption{$L_2$ errors compared to classical solutions. 
  Comparison of \textsc{prepare-select} (solid) and \textsc{arcsin} (hatched) encoding for 
  Toeplitz, l2d\_8x8\_ddrr, l2d\_4x4\_rrrr and l3d\_4x8x8\_dndddd cases.
  The condition numbers are the subnormalised values for each encoding.
  Phase factors for  $L_{\infty}=0.001$ and $\kappa_s>2000$ were not available.}
  \label{fig-ps-qo-e01-03-l2}
\end{figure}

\Cref{fig-ps-qo-e01-03-l2} compares the $L_2$ errors with \textsc{prepare-select} and
\textsc{arcsin} encoding for the Toeplitz, l2d\_4x4\_nnnn, l2d\_4x4\_rrrr 
and l3d\_4x8x8\_dndddd cases.
All the \textsc{prepare-select} cases use the $\kappa_s=50$ phase factors as used previously.
For \textsc{arcsin}, the 32x32 Toeplitz matrix with $\kappa=2.125$ gives an encoded
matrix with $\kappa_s=50$ allowing a direct comparison with \textsc{prepare-select}.
For the larger condition numbers, the longer run times of \textsc{qsppack} make it 
impractical to optimise the degree of the polynomial approximation.
For the \textsc{arcsin} cases, phase factors for subnormalised 
condition numbers 300, 1,000 and 2,500 were used.
Whilst this does not give an exact back to back comparison, it is indicative
of how QSVT would be used in practice. 
\textsc{qsppack} reliably generated all the phase factors used in this work
with the largest set having 16,813 factors.

%
\subsection{Success probability}
\label{subsec-success}

\begin{figure}[htb]
  \centering
  \captionsetup{justification=centering}
  \includegraphics[clip, trim=0.0cm 0.0cm 0.0cm 0.0cm,width=0.60\textwidth]{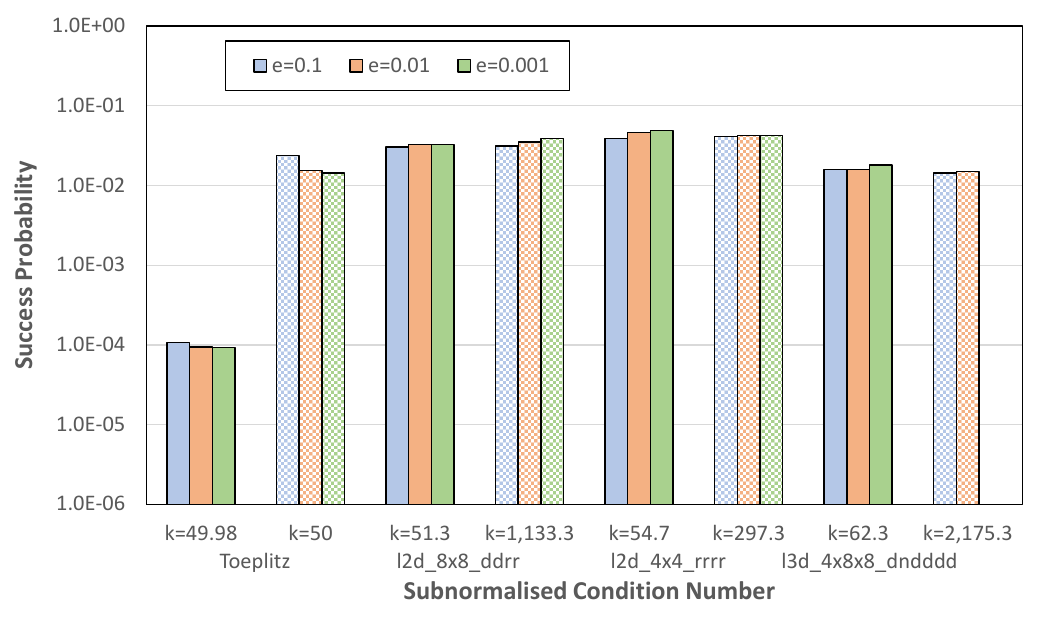}
  \caption{Success probabilities.
  Comparison of \textsc{prepare-select} (solid) and \textsc{arcsin} (hatched) encoding for 
  Toeplitz, l2d\_8x8\_ddrr, l2d\_4x4\_rrrr and l3d\_4x8x8\_dndddd cases.
  The condition numbers are the subnormalised values for each encoding.
  Phase factors for  $L_{\infty}=0.001$ and $\kappa>1000$ were not available.}
  \label{fig-k50-e01-e03-prob}
\end{figure}

\Cref{fig-k50-e01-e03-prob} shows the success probability of measuring the
signal and encoding qubits in the $\ket{0}$ state as in \Cref{fig-martyn_eval_BE},
using the same test cases as above.
Apart from the Toeplitz matrix with \textsc{prepare-select} encoding, the two
encoding methods have very similar success factors with small variations due
to the different $L_{\infty}$ errors levels of polynomial approximations.

Since the Toeplitz cases use different condition numbers to achieve
the same $\kappa_s$, the subnormalisation constants for
\textsc{prepare-select} and \textsc{arcsin} encoding are $s=3.34$ and
$s=32$ respectively. From \Cref{eqn-qsvt-expect}, the scaling 
with $s^2$ results in the 2 orders of magnitude
difference in the success probabilities.
For the other cases \Cref{eqn-qsvt-subnorm04} shows that, as 
long as $\kappa_s$ is chosen according to \Cref{eqn-qsvt-subnorm03},
the success probability is independent of the encoding method shown.

\begin{table}[htb]
  \centering
  \captionsetup{justification=centering}
  \begin{tabular}{c | c c | c c c c c }
    \toprule
    Measurement & Qubit & Expectation & Qubit & Expectation \\
    order & order & & order &\\
    \midrule
    $E_0$ &    9 & 9.583055e-01 &  16 & 6.105163e-02\\
    $E_1$ &   10 & 8.883451e-01 &   9 & 1.000000e+00\\
    $E_2$ &   11 & 9.170198e-01 &  10 & 1.000000e+00\\
    $E_3$ &   12 & 9.430148e-01 &  11 & 1.000000e+00\\
    $E_4$ &   13 & 9.522488e-01 &  12 & 1.000000e+00\\
    $E_5$ &   14 & 9.626014e-01 &  13 & 1.000000e+00\\
    $E_6$ &   15 & 9.754071e-01 &  14 & 1.000000e+00\\
    $E_7$ &   16 & 9.275374e-02 &  15 & 1.000000e+00\\
    \midrule
	E(Success) & & 6.105162e-02 & & 6.105162e-02 \\
    \bottomrule
    \\
  \end{tabular}
  \caption{Influence of the measurement order on the expectation of
           success for the l3d\_4x8x8\_dnrrdd test case. 
           Big endian ordering, qubit 16 = QSVT signal qubit.}
  \label{tab-esuccess-order}
\end{table}
 
An interesting observation on the expectations of the 
measurements is shown in \Cref{tab-esuccess-order}.
This is for the l3d\_4x8x8\_dnrrdd test case with \textsc{prepare-select}
encoding, but the same observation is true for \textsc{arcsin} encoding.
For this case, there are 9 qubits in the Select register,
7 qubits in the prepare register and one signal processing qubit.
With big endian ordering, the default is to measure the qubits in
ascending order, which measures the signal qubit last.
The flag qubits used in the encoding have expectations close
to one but not exactly one.
If, instead, the signal qubit is measured first and is in the 
$\ket{0}$ state, then all the flag qubits are also projected
onto the $\ket{0}$ state. At least, in the error-free 
simulations used herein.
As expected, the overall success probability is independent of
the measurement order.

%
\section{Results}
\label{sec-results}
These results focus on the use of QSVT as the linear
solver within an outer non-linear 
Computational Fluid Dynamics solver.
Two types of linear system are considered.
The first is the pressure-correction solver used within 
semi-implicit CFD schemes. 
The second is the coupled matrix from an implicit
CFD solver. Both are applied to the 2-dimensional lid-driven cavity
for which hybrid HHL solutions have been reported
\cite{lapworth2022hybrid,lapworth2022implicit}.
The same hybrid methodology has been used in this work.
Convergence plots show the change in the flow variables between the start
and end of each linear QSVT solution.
If the change is high, the non-linear set of equations are far from
being satisfied. As the outer non-linear iterations proceed, the
changes reduce towards zero at which point the non-linear solution
has been found.

All calculations were run on an
Intel\textsuperscript{\tiny\textregistered}
Core\textsuperscript{\tiny\textcopyright} i9 12900K 3.2GHz
Alder Lake 16 core processor with 64GB of 3,200MHz DDR4 RAM.

%
\subsection{Pressure correction equations}
\label{subsec-res-pc}

The smallest 2D pressure correction matrix is for a 5x5
CFD mesh that results in a 16x16 matrix.
From \Cref{tab-kappa-all}, this has subnormalised condition
numbers of 133 and 850 for \textsc{prepare-select} and \textsc{arcsin} encoding
respectively.

\begin{figure}[htb]
  \centering
  \captionsetup{justification=centering}
    \begin{subfigure}[b]{0.49\textwidth}
      \centering
      \includegraphics[width=.9\textwidth]{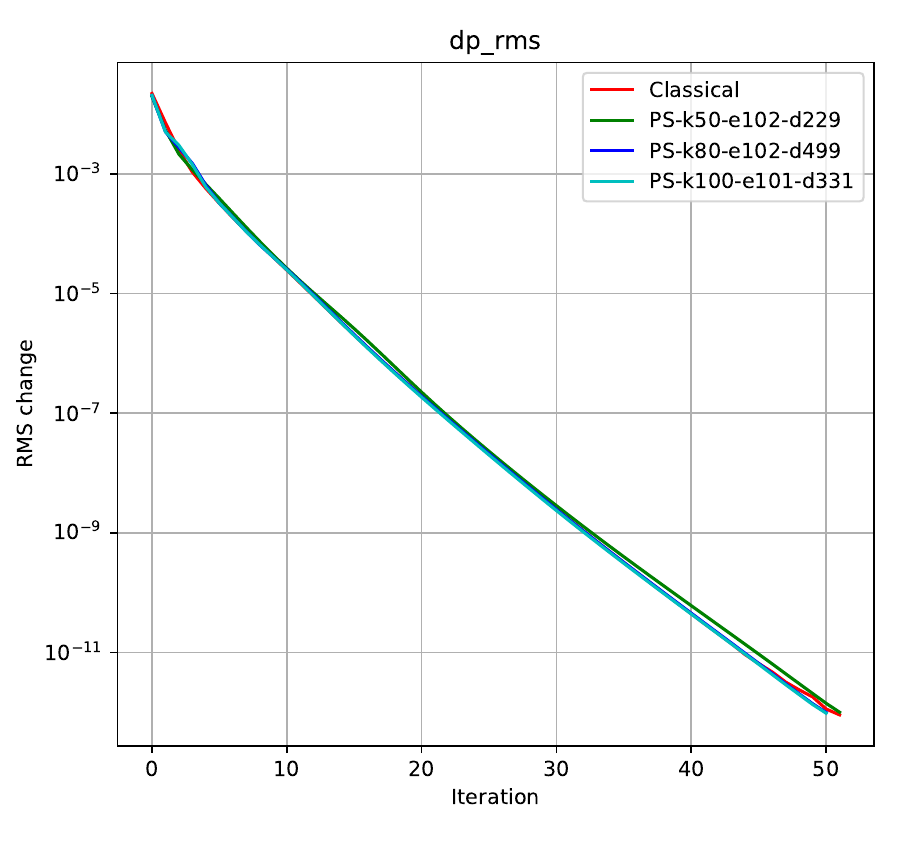}
      \caption{\textsc{prepare-select} encoding}
      \label{fig-cavity-4x4-pc-ps-dp}
  \end{subfigure}
  \captionsetup{justification=centering}
  \begin{subfigure}[b]{0.49\textwidth}
      \centering
      \includegraphics[width=.9\textwidth]{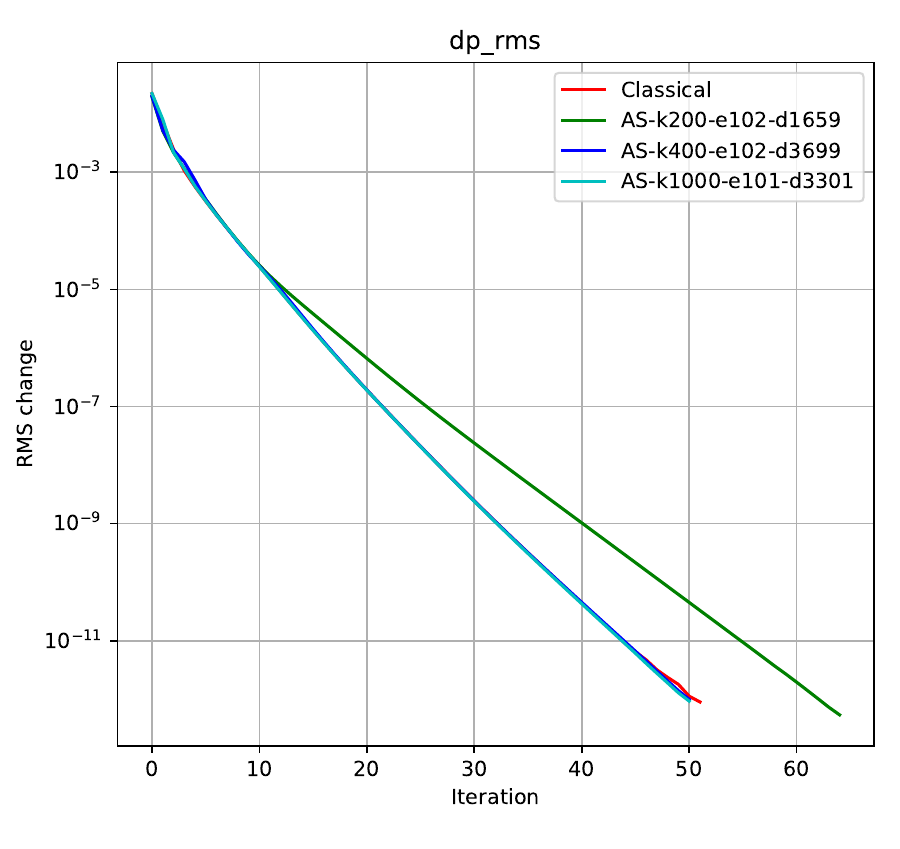}
      \caption{\textsc{arcsin} encoding.}
      \label{fig-cavity-4x4-pc-qo-dp}
  \end{subfigure}
  \caption{Convergence history of the pressure correction updates 
           for the  16x16 pressure correction using QSVT.
           k is conditional number, e is precision (e.g. 102 = $10^{-2}$),
           d is the number of phase factors.}
  \label{fig-cavity-4x4-pc-dp}
\end{figure}

\Cref{fig-cavity-4x4-pc-ps-dp} compares the classical solution with QSVT solutions
for $\kappa_s=50, 80, 100$ phase factors using \textsc{prepare-select} encoding. 
Even with $\kappa_s=50$ and $\epsilon=0.01$, the QSVT
convergence is almost identical to the classical one. These are well below the
expected condition number of 133.
\Cref{fig-cavity-4x4-pc-qo-dp} performs the same comparison for 
\textsc{arcsin} encoding where the expected condition number is 850.
Here, the $\kappa_s=1000, \epsilon=0.1$ and $\kappa_s=400, \epsilon=0.01$ phase factors
match the classical solution and have approximately the same number of
phase factors.
Even the $\kappa_s=200$ solution is reasonable.
Summed over all the non-linear iterations,
it has a total of 107,835 phase rotations compared to 168,351
for $\kappa_s=1000$.

\begin{figure}[htb]
  \centering
  \captionsetup{justification=centering}
    \begin{subfigure}[b]{0.49\textwidth}
      \centering
      \includegraphics[width=.9\textwidth]{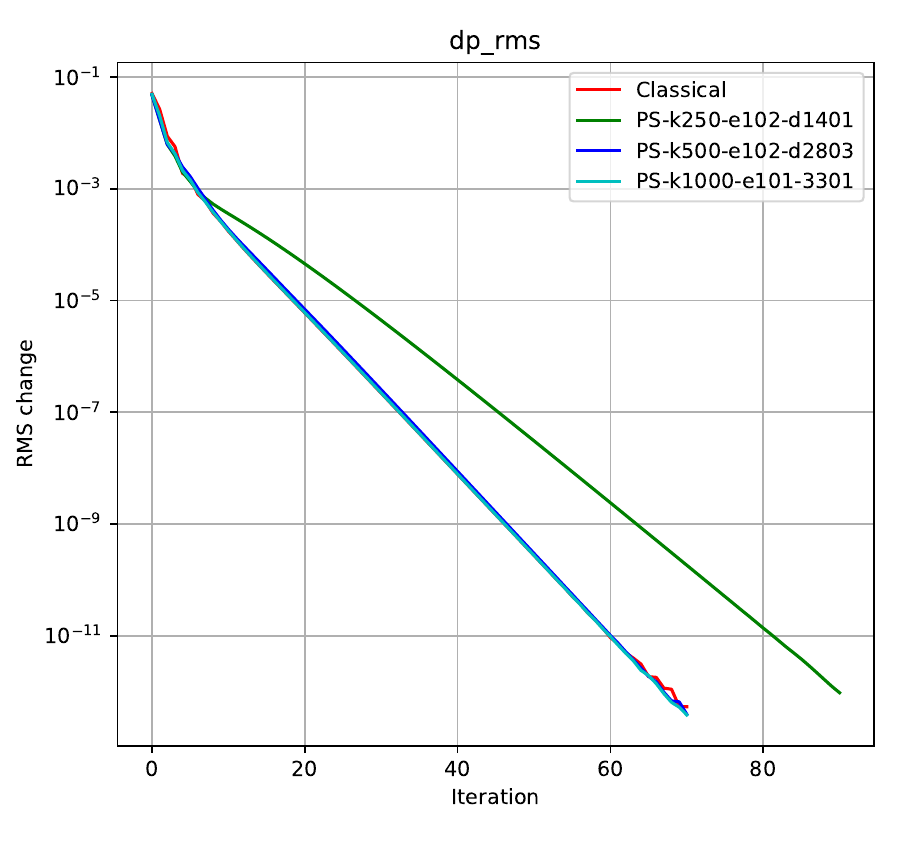}
      \caption{\textsc{prepare-select} encoding}
      \label{fig-cavity-8x8-pc-ps-dp}
  \end{subfigure}
  \captionsetup{justification=centering}
  \begin{subfigure}[b]{0.49\textwidth}
      \centering
      \includegraphics[width=.9\textwidth]{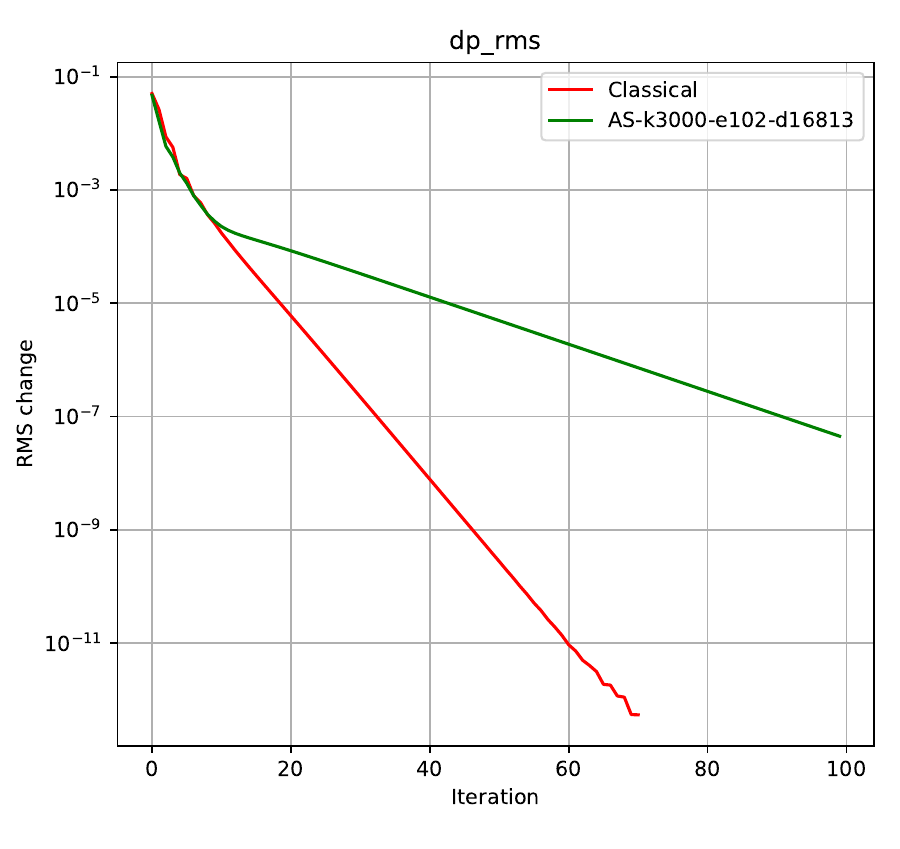}
      \caption{\textsc{arcsin} encoding.}
      \label{fig-cavity-8x8-pc-qo-dp}
  \end{subfigure}
  \caption{Convergence history of the pressure correction updates 
           for the  64x64 pressure correction using QSVT.
           k is conditional number, e is precision (e.g. 102 = $10^{-2}$),
           d is the number of phase factors.}
  \label{fig-cavity-8x8-pc-dp}
\end{figure}

\Cref{fig-cavity-8x8-pc-dp} compares the encoding techniques
on the 9x9 CFD mesh which has a 64x64 pressure correction matrix.
From \Cref{tab-kappa-all}, this has subnormalised condition
numbers of 1,186 and 22,063 for \textsc{prepare-select} and \textsc{arcsin} encoding
respectively.
The \textsc{prepare-select} results are similar to the results for the 16x16 
matrix.
For the \textsc{arcsin} results, the highest number of phase factors available 
was for $\kappa_s=3000, \epsilon=0.01$. 
It took \textsc{qsppack} 10.5 days to generate the 16,813 phase factors.
Most of this was applying the Remez algorithm to construct the
approximating polynomial \cite{dong2021efficient}. 
The BFGS optimisation to get the phase factors took 5.5 hours, which
could be reduced to less than an hour with Newton optimisation
\cite{dong2023robust}.
Assuming that phase factors 
for $\kappa_s=12000$ would be sufficient,
the $\mathcal{O}(\kappa^2)$ scaling of \textsc{qsppack} \cite{dong2021efficient}
means that it would take over 5 months to generate them.

\begin{figure}[htb]
  \centering
  \captionsetup{justification=centering}
  \includegraphics[clip, trim=0.0cm 0.0cm 0.0cm 0.0cm,width=0.60\textwidth]{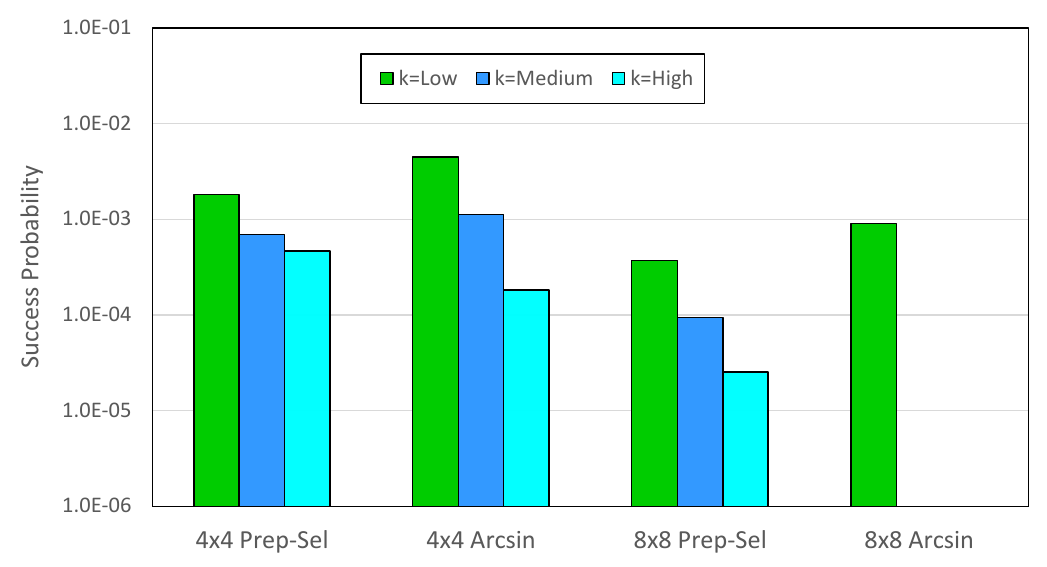}
  \caption{Success probabilities for pressure correction solver.
  Colours and order match \Cref{fig-cavity-4x4-pc-dp} and \Cref{fig-cavity-8x8-pc-dp}.}
  \label{fig-cavity-pc-prob}
\end{figure}

The QSVT results match previous findings for the effect of
precision on HHL \cite{lapworth2022implicit}.
The early iterations are dominated by high frequency error waves
corresponding to the largest eigenvalues. These are accurately
resolved by QSVT. However, these error waves are quickly eliminated
and low frequency waves, associated with the smaller eigenvalues, 
become dominant. As with HHL, failure to model the effects of
lowest eigenvalues leads to slow convergence rather than
divergence.

\Cref{fig-cavity-pc-prob} shows the success probabilities for matrices sampled
at 10 iterations from each of the above cases.
The change in expectation with condition number is consistent with
\Cref{eqn-qsvt-expect}.
For the 4x4 case using \textsc{arcsin} encoding, there is an extra
benefit of using $\kappa_s=200$ instead of $\kappa_s=1000$ as it gives
a factor 25 increase in success probability.

%
\subsection{Implicit coupled equations}
\label{subsec-res-cpl}
The smallest 2D implicit matrix is for a 5x5
CFD mesh that results in a 76x76 matrix which is padded with
an identity block to 128x128.
From \Cref{tab-kappa-all}, this has subnormalised condition
numbers of 331 and 900 for \textsc{prepare-select} and \textsc{arcsin} encoding
respectively.

\begin{figure}[htb]
  \centering
  \captionsetup{justification=centering}
    \begin{subfigure}[b]{0.49\textwidth}
      \centering
      \includegraphics[width=.9\textwidth]{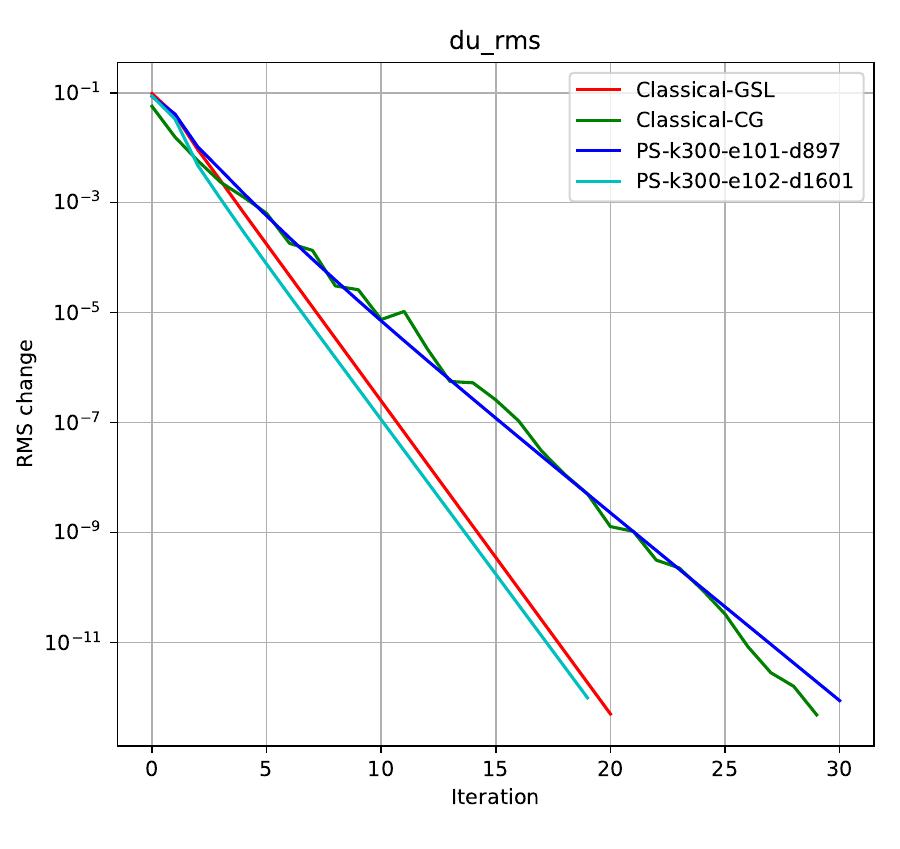}
      \caption{\textsc{prepare-select} encoding}
      \label{fig-cavity-5x5-cpl-ps-du}
  \end{subfigure}
  \captionsetup{justification=centering}
  \begin{subfigure}[b]{0.49\textwidth}
      \centering
      \includegraphics[width=.9\textwidth]{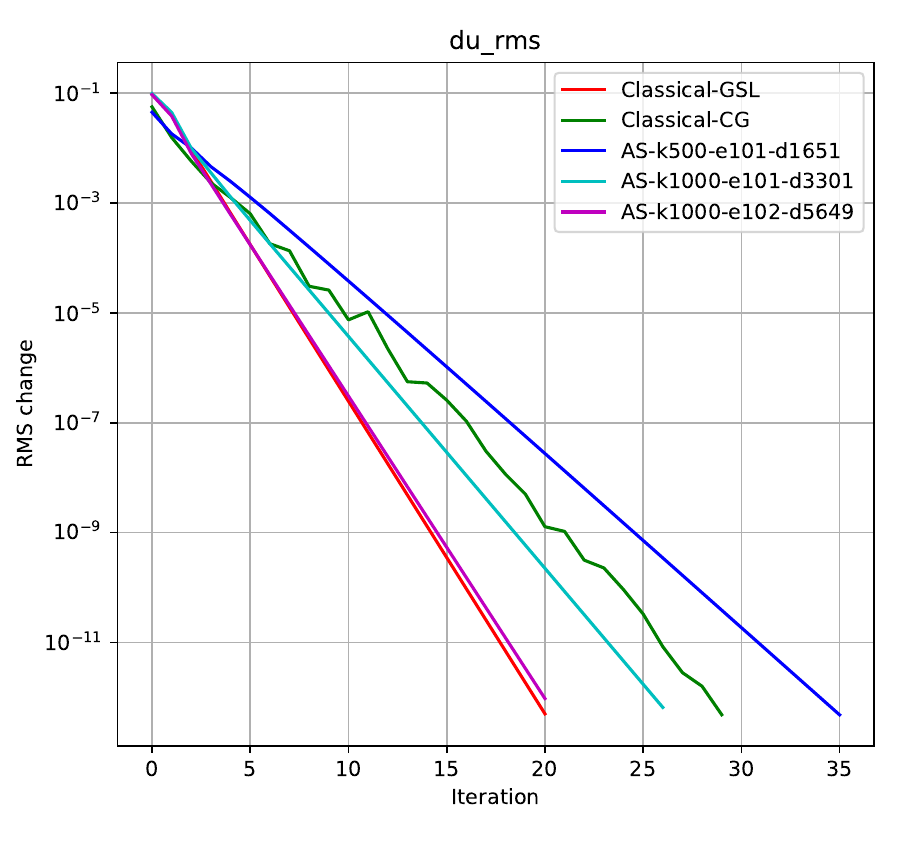}
      \caption{\textsc{arcsin} encoding.}
      \label{fig-cavity-5x5-cpl-qo-du}
  \end{subfigure}
  \caption{Convergence history of the velocity updates 
           for the 76x76 implicit matrix using QSVT.
           k is conditional number, e is precision (e.g. 102 = $10^{-2}$),
           d is the number of phase factors.}
  \label{fig-cavity-5x5-cpl-du}
\end{figure}

\Cref{fig-cavity-5x5-cpl-ps-du} compares the classical solution with 
QSVT solutions for $\kappa_s=300$ and $\epsilon=0.1, 0.02$ phase factors 
using \textsc{prepare-select} encoding. 
There are two classical solutions included in the plots.
The solution labelled 'GSL' uses an exact solver \cite{gough2009gnu} and
the solution labelled 'CG' uses an iterative conjugate gradient 
solver \cite{golub2013matrix}.
The lower precision QSVT solution matches the CG solution and the 
higher precision matches the GSL solution.
The \textsc{arcsin} solutions in \Cref{fig-cavity-5x5-cpl-qo-du}
show the same behaviour.
For the \textsc{arcsin} encoding, the pressure correction and implicit
matrices have similar condition numbers, $\kappa_s$, of around 900.
Whilst the pressure correction matrix can use phase factors for
$\kappa_s=400$ with no discernible effect, the implicit matrix has a
near doubling of the number of non-linear iterations if 
phase factors for $\kappa_s=500$ are used.

\begin{figure}[htb]
  \centering
  \captionsetup{justification=centering}
    \begin{subfigure}[b]{0.49\textwidth}
      \centering
      \includegraphics[width=.9\textwidth]{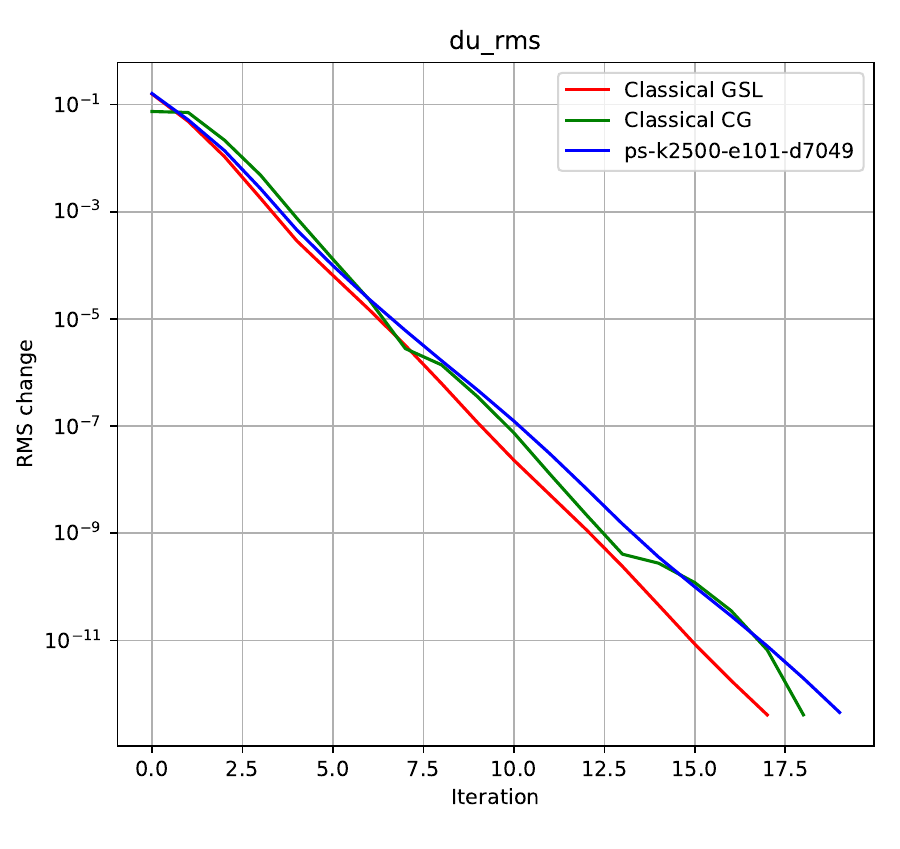}
      \caption{\textsc{prepare-select} encoding}
      \label{fig-cavity-9x9-cpl-ps-du}
  \end{subfigure}
  \captionsetup{justification=centering}
  \begin{subfigure}[b]{0.49\textwidth}
      \centering
      \includegraphics[width=.9\textwidth]{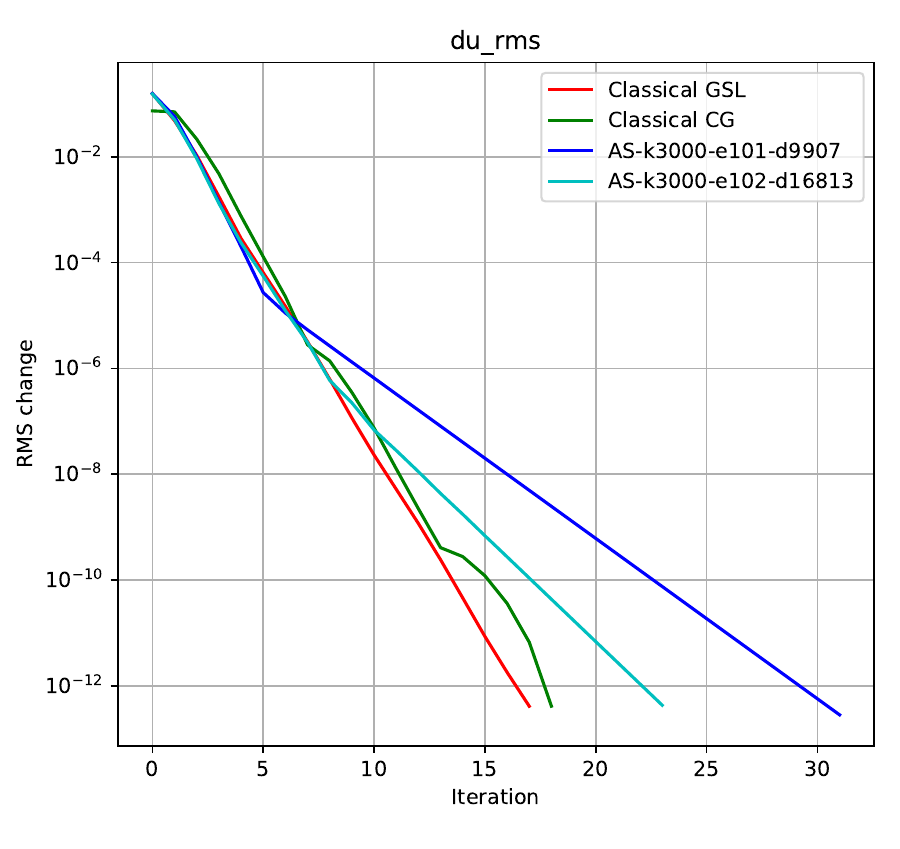}
      \caption{\textsc{arcsin} encoding.}
      \label{fig-cavity-9x9-cpl-qo-du}
  \end{subfigure}
  \caption{Convergence history of the velocity updates 
           for the 244x244 implicit matrix using QSVT.
           Labelling as before.}
  \label{fig-cavity-9x9-cpl-du}
\end{figure}

\Cref{fig-cavity-9x9-cpl-ps-du} compares the classical and QSVT
coupled solutions for the 9x9 CFD mesh with \textsc{prepare-select}
encoding, for which $\kappa_s=2,407$.
Given the long run times for this case, see \Cref{subsubsec-comptime}, only one QSVT
calculation was performed with phase factors for
$\kappa_s=2,500, \epsilon=0.1$. The QSVT solution matches the
Classical conjugate gradient solution.
Note that the CG and GSL classical solutions are more similar but
the CG solver required a single SIMPLE iteration at the start to 
prevent it diverging. The QSVT solution did not need this.
A further indication that QSVT may yield stability benefits.
\Cref{fig-cavity-9x9-cpl-qo-du} compares the classical and QSVT
coupled solutions with \textsc{arcsin} encoding, for
which $\kappa_s=2,824$. 
Due to the time taken by \textsc{qsppack} to generate phase factors
only two QSVT solutions are available with
$\kappa_s=3,000$ and $\epsilon=0.01, 0.1$.
These show the same trends as for the 5x5 coupled calculations.

\begin{figure}[htb]
  \centering
  \captionsetup{justification=centering}
  \includegraphics[clip, trim=0.0cm 0.0cm 0.0cm 0.0cm,width=0.60\textwidth]{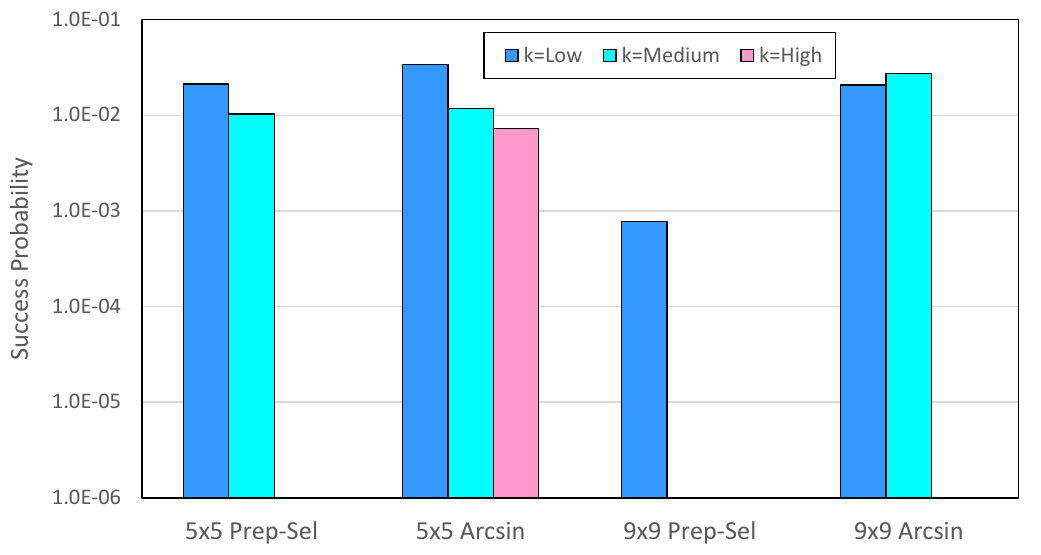}
  \caption{Success probabilities for implicit solver.
  Colours and order match \Cref{fig-cavity-5x5-cpl-du} and \Cref{fig-cavity-9x9-cpl-du}.}
  \label{fig-cavity-cpl-prob}
\end{figure}

\Cref{fig-cavity-cpl-prob} shows the success probabilities for coupled
QSVT solutions. The scaling with $\kappa_s$ shows the expected behaviour.
Of note is that the coupled solver has higher success probabilities than
the pressure correction solver. 
This is in part due to lower condition numbers, but also due to the fact
that the pressure correction matrices have to be scaled down to 
achieve $|A|_{max}=1$ whereas the coupled matrices can be scaled up.
These differences are included in the values in \Cref{tab-kappa-all}.

%
\subsubsection{Computational considerations}
\label{subsubsec-comptime}

For the \textsc{prepare-select} encoding of the 9x9 coupled case, 
the LCU contained 16,104 Pauli strings although the time to 
compute these was less than a minute.
The main computational consideration was that the LCU resulted in 
the \textsc{prepare} register having
14 qubits compared to 8 in the equivalent \textsc{row} register.
The \textsc{prepare} operations are the most time consuming to emulate
and, even with the optimisations described in \Cref{app-subsec-emul-ps},
each phase factor step took 2.4s, compared to 0.01s for \textsc{arcsin}.
For the \textsc{prepare-select} result in \Cref{fig-cavity-9x9-cpl-ps-du}
with 7,049 phase factors,
each complete QSVT solve took just under 5 hours compared to less than 
3 minutes for \textsc{arcsin} with 16,183 phase factors.
Whilst classical emulation times should not be taken as indicative
of the time on a physical quantum computer, the difference in
the operational intensity of the two encoding methods is likely
to have some relevance.

%
\section{Conclusions}
\label{sec-conclude}

A sparse matrix query oracle using \textsc{arcsin} encoding generates
circuits that scale with the number of non-zeros in the matrix.
This has minimal classical preprocessing costs and addresses one of the
major issues of \textsc{prepare-select} encoding based on Pauli 
strings - the classical preprocessing time to construct the LCU.
The downside of the \textsc{arcsin} encoding is that the subnormalisation
constant $s$ scales with the size of the matrix. 
For pressure correction based CFD solvers this leads
to an $\mathcal{O}(n_d/N)$ overhead relative to \textsc{prepare-select}
encoding, where  $n_d$ is the number of diagonals in the matrix.
However, for the implicit coupled CFD solver, the vastly increased
number of Pauli strings in the LCU gives greater parity in the
subnormalisation constants.

A key part of QSVT is setting $\kappa_s$ to accurately resolve the
influence of the eigenvectors with the lowest eigenvalues. These are
responsible for long-wavelength errors in the CFD flow field and setting
$\kappa_s$ too low slows the rate of convergence but does not, generally, 
lead to divergence. Given that lowering $\kappa_s$ increases the 
success probability, optimising the performance of QSVT within a
CFD code is a function of the number QSVT phase factors, 
the number of non-linear iterations and the number of shots.

The $L_{\infty}$ approximation errors in the QSVT phase factors
has been found to less impactful than the condition number.
Error tolerances of $10^{-2}$ were more than sufficient in most cases
and in some cases $10^{-1}$ was sufficient. There are some indications
that QSVT is more robust in the first iterations of an implicit solver 
that often require special treatment in classical solvers. 
However, the cases are orders of magnitude from industrial matrices 
results and these inferences must be treated with caution.

With the goal of minimising the time spent on a classical computer 
using hybrid implicit CFD solvers, \textsc{arcsin} encoding for the
\textsc{query-oracle} has shown that QSVT can be an effective
linear solver. However, the time taken to generate the
QSVT phase factors remains a significant impediment to scaling QSVT
to larger test cases.
Unlike constructing a LCU, the calculation of each set of
phase factors is a one-off calculation separate from the CFD solver.
The unifying nature of QSVT means that a library of phase factors
can be used to invert a wide range of matrix types.

%
\section{Data availability}
\label{sec-data}
The L-QLES input files for the Laplacian test cases are available from
\href{https://github.com/rolls-royce/qc-cfd}{https://github.com/rolls-royce/qc-cfd}.

%
\section{Acknowledgements}
I would like to thank Christoph S{\"u}nderhauf of Riverlane 
for his guidance and many useful discussions on the implementation of QSVT.
I would also like to thank Bjorn Berntson and Zal Nemeth of Riverlane for
their help in choosing the input parameters for \textsc{QSPPACK}.
I would also like to thank Jarrett Smalley and Tony Phipps 
of Rolls-Royce for their helpful comments on this work.
This work would not have been possible without \textsc{QSPPACK} and I would like
to thank Yulong Dong of UC Berkeley for his advice at the outset of this
work.

The permission of Rolls-Royce to publish this work is gratefully acknowledged.
This work was completed under funding received under the UK's
Commercialising Quantum Technologies Programme (Grant reference 10004857).

%
\newpage
\bibliographystyle{ieeetr}
\bibliography{references}

\begin{thebibliography}{10}

\bibitem{uk_quantum_missions}
D.~for Science~Innovation and Technology, ``National quantum strategy missions.'' \url{https://www.gov.uk/government/publications/national-quantum-strategy/national-quantum-strategy-missions}, 2023.
\newblock Accessed: 2024-02-15.

\bibitem{lapworth2022hybrid}
L.~Lapworth, ``A hybrid quantum-classical cfd methodology with benchmark hhl solutions,'' {\em arXiv preprint arXiv:2206.00419}, 2022.

\bibitem{lapworth2022implicit}
L.~Lapworth, ``Implicit hybrid quantum-classical cfd calculations using the hhl algorithm,'' {\em arXiv preprint arXiv:2209.07964}, 2022.

\bibitem{harrow2009quantum}
A.~W. Harrow, A.~Hassidim, and S.~Lloyd, ``Quantum algorithm for linear systems of equations,'' {\em Physical review letters}, vol.~103, no.~15, p.~150502, 2009.

\bibitem{childs2012hamiltonian}
A.~M. Childs and N.~Wiebe, ``Hamiltonian simulation using linear combinations of unitary operations,'' {\em arXiv preprint arXiv:1202.5822}, 2012.

\bibitem{trotter1959product}
H.~F. Trotter, ``On the product of semi-groups of operators,'' {\em Proceedings of the American Mathematical Society}, vol.~10, no.~4, pp.~545--551, 1959.

\bibitem{babbush2018encoding}
R.~Babbush, C.~Gidney, D.~W. Berry, N.~Wiebe, J.~McClean, A.~Paler, A.~Fowler, and H.~Neven, ``Encoding electronic spectra in quantum circuits with linear t complexity,'' {\em Physical Review X}, vol.~8, no.~4, p.~041015, 2018.

\bibitem{berry2015simulating}
D.~W. Berry, A.~M. Childs, R.~Cleve, R.~Kothari, and R.~D. Somma, ``Simulating hamiltonian dynamics with a truncated taylor series,'' {\em Physical review letters}, vol.~114, no.~9, p.~090502, 2015.

\bibitem{Low2019hamiltonian}
G.~H. Low and I.~L. Chuang, ``Hamiltonian {S}imulation by {Q}ubitization,'' {\em {Quantum}}, vol.~3, p.~163, July 2019.

\bibitem{wiebe2021theory}
N.~Wiebe and S.~Zhu, ``A theory of trotter error,'' {\em Physical Review X}, vol.~11, no.~011020, p.~26, 2021.

\bibitem{martyn2021grand}
J.~M. Martyn, Z.~M. Rossi, A.~K. Tan, and I.~L. Chuang, ``Grand unification of quantum algorithms,'' {\em PRX Quantum}, vol.~2, p.~040203, Dec 2021.

\bibitem{gilyen2019quantum}
A.~Gily{\'e}n, Y.~Su, G.~H. Low, and N.~Wiebe, ``Quantum singular value transformation and beyond: exponential improvements for quantum matrix arithmetics,'' in {\em Proceedings of the 51st Annual ACM SIGACT Symposium on Theory of Computing}, pp.~193--204, 2019.

\bibitem{dong2021efficient}
Y.~Dong, X.~Meng, K.~B. Whaley, and L.~Lin, ``Efficient phase-factor evaluation in quantum signal processing,'' {\em Physical Review A}, vol.~103, no.~4, p.~042419, 2021.

\bibitem{lin2022lecture}
L.~Lin, ``Lecture notes on quantum algorithms for scientific computation,'' {\em arXiv preprint arXiv:2201.08309}, 2022.

\bibitem{clader2022quantum}
B.~D. Clader, A.~M. Dalzell, N.~Stamatopoulos, G.~Salton, M.~Berta, and W.~J. Zeng, ``Quantum resources required to block-encode a matrix of classical data,'' {\em IEEE Transactions on Quantum Engineering}, vol.~3, pp.~1--23, 2022.

\bibitem{childs2017quantum}
A.~M. Childs, R.~Kothari, and R.~D. Somma, ``Quantum algorithm for systems of linear equations with exponentially improved dependence on precision,'' {\em SIAM Journal on Computing}, vol.~46, no.~6, pp.~1920--1950, 2017.

\bibitem{chakraborty2018power}
S.~Chakraborty, A.~Gily{\'e}n, and S.~Jeffery, ``The power of block-encoded matrix powers: improved regression techniques via faster hamiltonian simulation,'' {\em arXiv preprint arXiv:1804.01973}, 2018.

\bibitem{camps2022explicit}
D.~Camps, L.~Lin, R.~Van~Beeumen, and C.~Yang, ``Explicit quantum circuits for block encodings of certain sparse matrice,'' {\em arXiv preprint arXiv:2203.10236}, 2022.

\bibitem{motlagh2023generalized}
D.~Motlagh and N.~Wiebe, ``Generalized quantum signal processing,'' {\em arXiv preprint arXiv:2308.01501}, 2023.

\bibitem{camps2022fable}
D.~Camps and R.~Van~Beeumen, ``Fable: Fast approximate quantum circuits for block-encodings,'' in {\em 2022 IEEE International Conference on Quantum Computing and Engineering (QCE)}, pp.~104--113, IEEE, 2022.

\bibitem{sunderhauf2023block}
C.~S{\"u}nderhauf, E.~Campbell, and J.~Camps, ``Block-encoding structured matrices for data input in quantum computing,'' {\em arXiv preprint arXiv:2302.10949}, 2023.

\bibitem{gough2009gnu}
B.~Gough, {\em GNU scientific library reference manual}.
\newblock Network Theory Ltd., 2009.

\bibitem{lapworth2024qles}
L.~Lapworth, ``L-qles: Sparse laplacian generator for evaluating quantum linear equation solvers,'' {\em arXiv preprint arXiv:2402.12266}, 2024.

\bibitem{kothari2014efficient}
R.~Kothari, {\em Efficient algorithms in quantum query complexity}.
\newblock PhD thesis, University of Waterloo, 2014.

\bibitem{berry2018improved}
D.~W. Berry, M.~Kieferov{\'a}, A.~Scherer, Y.~R. Sanders, G.~H. Low, N.~Wiebe, C.~Gidney, and R.~Babbush, ``Improved techniques for preparing eigenstates of fermionic hamiltonians,'' {\em npj Quantum Information}, vol.~4, no.~1, pp.~1--7, 2018.

\bibitem{mottonen2004transformation}
M.~Mottonen, J.~J. Vartiainen, V.~Bergholm, and M.~M. Salomaa, ``Transformation of quantum states using uniformly controlled rotations,'' {\em arXiv preprint quant-ph/0407010}, 2004.

\bibitem{araujo2021divide}
I.~F. Araujo, D.~K. Park, F.~Petruccione, and A.~J. da~Silva, ``A divide-and-conquer algorithm for quantum state preparation,'' {\em Scientific Reports}, vol.~11, no.~1, pp.~1--12, 2021.

\bibitem{shende2004synthesis}
V.~V. Shende, S.~S. Bullock, and I.~L. Markov, ``Synthesis of quantum logic circuits,'' {\em arXiv preprint quant-ph/0406176}, 2004.

\bibitem{shende2005synthesis}
V.~V. Shende, S.~S. Bullock, and I.~L. Markov, ``Synthesis of quantum logic circuits,'' in {\em Proceedings of the 2005 Asia and South Pacific Design Automation Conference}, pp.~272--275, 2005.

\bibitem{gilyen2019phd}
A.~Gily{\'e}n, {\em Quantum singular value transformation \& its algorithmic applications}.
\newblock PhD thesis, University of Amsterdam, 2019.

\bibitem{dong2023robust}
Y.~Dong, L.~Lin, H.~Ni, and J.~Wang, ``Robust iterative method for symmetric quantum signal processing in all parameter regimes,'' {\em arXiv preprint arXiv:2307.12468}, 2023.

\bibitem{golub2013matrix}
G.~H. Golub and C.~F. Van~Loan, {\em Matrix computations}.
\newblock JHU press, 2013.

\end{thebibliography}

%
\newpage
\appendix
%
\section{Test case details}
\label{app-sec-cases}

In the following tables,
the naming convention for the Laplacian cases, those with 'l' as the
first character is: \texttt{dimension\_mesh\_bcs}.
The input files for these cases are available at 
\href{https://github.com/rolls-royce/qc-cfd}{github.com/rolls-royce/qc-cfd}.
For CFD matrices labelled \texttt{cavity-pc}, the mesh dimensions are for the
pressure correction mesh.
For the \texttt{cavity-cpl} matrices, the mesh dimensions are for the nodal
mesh. For example, \texttt{cavity-pc-4x4} and \texttt{cavity-cpl-5x5}
are different solvers on the same mesh.

For the coupled CFD cases, the matrix dimension is that of the original
CFD system. As described in \cite{lapworth2022implicit}, the matrix is
padded with a diagonal block to give a dimension that is the next
largest power of 2.

\subsection{Condition numbers}
\label{app-subsec-kappa}

\begin{table}[h]
  \centering
    \captionsetup{justification=centering}
  \begin{tabular}{l l c c c c c c }
    \toprule
Index & Case name & rank (A) & $\kappa$ (A) & $\kappa$ (AH)  &  $\kappa$ (PS) & $\kappa$ (QO) \\
    \midrule
1  & l1d\_8\_dd	         &	8	 &	15.3	& 16.4	&	26.1	&	84.7	\\
2  & l1d\_8\_rr	         &	8	 &	25.3    & 26.0	&	41.8	&	105.1	\\
3  & l1d\_16\_dd	         &	16	 &	99.1	& 107.3	&	180.9	&	1,134.5	\\
4  & l1d\_32\_dd	         &	32	 &	698.3	& 735.8	&	1287.7	&	15,101.5	\\
\\
5  & l2d\_4x4\_dddd	     &	16	 &	2.4 	& 2.9	&	3.8 	&	27.4	\\
6  & l2d\_4x4\_nnnn	     &	16	 &	25.7	& 41.3	&	77.2	&	213.8	\\
7  & l2d\_4x4\_rrrr	     &	16	 &	36.2	& 37.3	&	54.7	&	297.3	\\
8  & l2d\_8x8\_dddd	     &	64	 &	13.5	& 13.8	&	27.5	&	544.0	\\
9  & l2d\_8x8\_ddrr	     &	64	 &	33.1	& 33.3	&	51.3	&	1,133.3	\\
10 & l2d\_8x8\_nnnn	     &	64	 &	155.8	& 189.1	&	527.3	&	5,583.2	\\
11 & l2d\_8x8\_rrrr	     &	64	 &	199.5	& 200.4	&	346.9	&	6,416.1	\\
12 & l2d\_16x16\_dddd	 &	256	 &	62.0    & 67.0  &	196.7	&	12,151.6	\\
13 & l2d\_16x16\_ddrr	 &	256	 &	140.7   & 142.0	&	317.8	&	23,956.2	\\
14 & l2d\_32x32\_dddd	 &	1,024 &	401.3	& 423.6	&	1,501.5	&	298,501.1	\\
\\
15 & l3d\_4x4x4\_dndddd	 &	64	 &	3.2	    & 5.5	&	5.9 	&	113.2	\\
16 & l3d\_4x8x8\_dndddd	 &	256	 &	14.4	& 22.8	&	62.3	&	2,175.3	\\
17 & l3d\_4x8x8\_dnrrdd	 &	256	 &	32.6    & 50.0	&	100.5	&	4,405.3	\\
18 & l3d\_8x8x8\_dddddd	 &	512	 &	14.5	& 14.7	&	37.1	&	4,127.9	\\
19 & l3d\_8x8x8\_ddrrdd	 &	512	 &	24.6	& 24.7	&	48.0	&	6,752.8	\\
20 & l3d\_8x8x8\_dnrrdd	 &	512	 &	44.7	& 52.3	&	121.1	&	11,897.0	\\
21 & l3d\_8x16x16\_dndddd &	2,048 &	62.4	& 93.1	&	422.8	&	95,420.9	\\
22 & l3d\_8x16x16\_dnrrdd &	2,048 &	136.8	& 183.6	&	615.2	&	181,163.7	\\
23 &l3d\_16x16x16\_dddddd &	4,096 &	82.7	& 98.1	&	301.7	&	171,397.3	\\
\\
24 & cavity-pc-4x4	     &	16	 &	87.7	&	88.7	&	133.5	&	851.2	\\
25 & cavity-pc-8x8	     &	64	 &	567.3	&	568.3	&	1,186.2 &	22,063.6	\\
26 & cavity-cpl-5x5       &	76	 &	18.3	&	20.9	&	331.4	&	900.3	\\
27 & cavity-cpl-6x6	     &	109	 &	28.2	&	28.2	&	583.4	&	1,021.1	\\
28 & cavity-cpl-9x9	     &	244	 &	30.5	&	72.8	&	2,407.4	&	2,823.5	\\
29 & cavity-cpl-13x13     &	508	 &	105.7	&	205.9	&	11,707.9	&	24,374.5	\\
30 & cavity-cpl-17x17     &	868	 &	275.5	&	483.8	&	-	&	156,973.6	\\
    \bottomrule
    \\
  \end{tabular}
  \caption{Full matrix rank and condition numbers $\kappa$ for each case.
  A = original non-Hermitian matrix, AH = symmetrised Hermitian matrix,
  PS = \textsc{prepare-select} encoding using AH, 
  QO = Query Oracle (\textsc{arcsin} and \textsc{fable}) encoding using A.}
  \label{tab-kappa-all}
\end{table}

\Cref{tab-kappa-all} gives the condition numbers for all the cases
used in this study. The condition numbers are computed using the 
GNU Scientific Library \cite{gough2009gnu}.
There are 4 columns of condition number:
\begin{itemize}
    \item $\kappa$ (A) - the original non-Hermitian matrices.
    \item $\kappa$ (AH) - the symmetrised Hermitian matrices.
\begin{equation}
  AH = 
  \begin{pmatrix}
    0           & A \\
    A^{\dagger} & 0
  \end{pmatrix}
  \label{eqn-AH}
\end{equation}
    \item $\kappa$ (PS) - the sub-normalised matrices using \textsc{prepare-select}
    encoding. The condition number for the 17x17 coupled matrix is absent
    as the time to compute the LCU was too excessive.
    \item $\kappa$ (QO) - the sub-normalised matrices using \textsc{arcsin} 
    and \textsc{fable} encoding.
\end{itemize}

\subsection{Operation counts}
\label{app-subsec-nopers}

\begin{table}[h]
  \centering
    \captionsetup{justification=centering}
  \begin{tabular}{l l c c c c c c }
    \toprule
    Index & Case name        & rank (A) & \#non-zeros (A)  & \textsc{prepare-select}  &  \textsc{fable} & \textsc{arcsin} \\
    \midrule
 1 & l1d\_8\_dd	         &	8	 &	20	& 42	& 32	& 20	\\
 2 & l1d\_8\_rr	         &	8	 &	22	& 57	& 42	& 18	\\
 3 & l1d\_16\_dd	         &	16	 &  44	& 96	& 128	& 43	\\
 4 & l1d\_32\_dd	         &	32	 &  92	& 216	& 512	& 88	\\
\\
 5 & l2d\_4x4\_dddd	     &	16	 &	32	& 60	& 64	& 32	\\
 6 & l2d\_4x4\_nnnn	     &	16	 &	40	& 204	& 208	& 40	\\
 7 & l2d\_4x4\_rrrr	     &	16	 &	76	& 195	& 166	& 53	\\
 8 & l2d\_8x8\_dddd	     &	64	 &	208	& 288	& 1,024	& 208	\\
 9 & l2d\_8x8\_ddrr	     &	64	 &	256	& 102	& 384	& 232	\\
10 & l2d\_8x8\_nnnn	     &	64	 &	232	& 864	& 3,328	& 232	\\
11 & l2d\_8x8\_rrrr	     &	64	 &	316	& 783	& 2,608	& 252	\\
12 & l2d\_16x16\_dddd	 &	256	 &	1,040 &	1,344 & 16,104  &	990	\\
13 & l2d\_16x16\_ddrr	 &	256	 &	1,152 &	360	  & 5,632	& 1,013	\\
14 & l2d\_32x32\_dddd	 &	1,024 &	4,624 &	6,141 &	188,825	& 4,384	\\
\\
15 & l3d\_4x4x4\_dndddd	 &	64	 &	116    & 180	 & 1,024	& 112	\\
16 & l3d\_4x8x8\_dndddd	 &	256	 &	724	   & 816 & 16,332	& 686	\\
17 & l3d\_4x8x8\_dnrrdd	 &	256	 &	880	   & 264 & 5,116	& 777	\\
18 & l3d\_8x8x8\_dddddd	 &	512	 &	1,808  & 1,629 & 31,919	& 1,808	\\
19 & l3d\_8x8x8\_ddrrdd	 &	512	 &	2,240  & 528 & 12,192	& 2,096	\\
20 & l3d\_8x8x8\_dnrrdd	 &	512	 &	2,288  & 540 & 20,477	& 2,288	\\
21 & l3d\_8x16x16\_dndddd &	2,048 &	9,300  & 7,296 & 589,491 & 8,183	\\
22 & l3d\_8x16x16\_dnrrdd &	2,048 &	10,336 & 1,704  & 198,452 & 8,588	\\
23 & l3d\_16x16x16\_dddddd &	4,096 & 9,104  & 6,147 & 292,884 & 7,144	\\
\\
24 & cavity-pc-4x4	     &	16	 &	62	  & 189	    & 256	   & 62	\\
25 & cavity-pc-8x8	     &	64	 &	286	  & 957	    & 4,074	   & 286	\\
26 & cavity-cpl-5x5       &	76	 &	317	  & 13,536  & 16,377	   & 308	\\
27 & cavity-cpl-6x6	     &	109	 &	440	  & 15,696  & 16,354	   & 440	\\
28 & cavity-cpl-9x9	     &	244	 &	1,117 & 48,312  & 65,401	   & 1,103	\\
29 & cavity-cpl-13x13     &	508	 &	2,525 & 146,868 & 259,326   & 2,525	\\
30 & cavity-cpl-17x17     &	868	 &	4,669 & -	    & 1,003,800 & 4,608	\\
    \bottomrule
    \\
  \end{tabular}
  \caption{Number of operations for each test case and encoding scheme.}
  \label{tab-nopers-all}
\end{table}

In \Cref{tab-nopers-all} the number of operations are counted as follows.
For both \textsc{arcsin} and \textsc{fable} encoding, the number of rotation gates is used.
This ignores the fact that \textsc{arcsin} uses multi-controlled rotations and
\textsc{fable} uses single qubit rotations. The number of CNOT gates in \textsc{fable} circuit
is also ignored.
For \textsc{prepare-select}, the number of terms in the LCU is used. This
is multiplied by a factor of 3 as the \textsc{prepare} circuit, its adjoint
and the Select circuit have the same number of operations.
Hadamard and swap gates that scale with the number of qubits are
not counted.
%
\section{Encoding a 2x2 matrix encoding}
\label{app-2x2enc}

\begin{figure}[ht]
  \centering
  \includegraphics[width=0.80\textwidth]{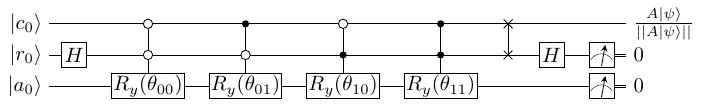}
  \caption{Circuit for full encoding of a 2x2 matrix using big-endian indexing.}
  \label{fig-app_2x2_orig_big}
\end{figure}

This section shows the assembly of the query oracle encoding
of a 2x2 matrix as this may be of interest to some readers.
The encoding circuit is shown in \Cref{fig-app_2x2_orig_big}. 

Each of the controlled rotations in \Cref{fig-app_2x2_orig_big} can be written as:
\begin{equation}
    U_{ij} = I^{\otimes 3} + (R_{ij}-I) \otimes M_i \otimes M_j
\end{equation}

Where $R_{ij}$ is shorthand for $R(\theta_{ij})$ and $i,j \in {0,1}$.
The $M$ matrices are the single qubit measurement operators in the 
computational basis:

\begin{equation}
  M_0 = \ket{0}\bra{0} =
  \begin{pmatrix}
    1 & 0 \\
    0 & 0
  \end{pmatrix}
  \label{eqn-M0}
\end{equation}

\begin{equation}
  M_1 = \ket{1}\bra{1} =
  \begin{pmatrix}
    0 & 0 \\
    0 & 1
  \end{pmatrix}
  \label{eqn-M1}
\end{equation}

Expanding for $U_{00}$ gives:

\begin{equation}
  U_{00} = 
  \begin{pmatrix}
    1 & 0 & 0 & 0 & 0 & 0 & 0 & 0\\
    0 & 1 & 0 & 0 & 0 & 0 & 0 & 0\\
    0 & 0 & 1 & 0 & 0 & 0 & 0 & 0\\
    0 & 0 & 0 & 1 & 0 & 0 & 0 & 0\\
    0 & 0 & 0 & 0 & 1 & 0 & 0 & 0\\
    0 & 0 & 0 & 0 & 0 & 1 & 0 & 0\\
    0 & 0 & 0 & 0 & 0 & 0 & 1 & 0\\
    0 & 0 & 0 & 0 & 0 & 0 & 0 & 0\\
  \end{pmatrix}
  +
  \begin{pmatrix}
     c_{00}-1 & -s_{00} \\
     s_{00}   &  c_{00} -1
  \end{pmatrix}
  \otimes
    \begin{pmatrix}
    1 & 0 & 0 & 0 \\
    0 & 0 & 0 & 0 \\
    0 & 0 & 0 & 0 \\
    0 & 0 & 0 & 0 \\
  \end{pmatrix}
  \label{eqn-U00-1}
\end{equation}

Giving

\begin{equation}
  U_{00} = 
  \begin{pmatrix}
    c_{00} & 0 & 0 & 0 & -s_{00} & 0 & 0 & 0\\
    0      & 1 & 0 & 0 & 0       & 0 & 0 & 0\\
    0      & 0 & 1 & 0 & 0       & 0 & 0 & 0\\
    0      & 0 & 0 & 1 & 0       & 0 & 0 & 0\\
    s_{00} & 0 & 0 & 0 & c_{00}  & 0 & 0 & 0\\
    0      & 0 & 0 & 0 & 0       & 1 & 0 & 0\\
    0      & 0 & 0 & 0 & 0       & 0 & 1 & 0\\
    0      & 0 & 0 & 0 & 0       & 0 & 0 & 0\\
  \end{pmatrix}
  \label{eqn-U00-2}
\end{equation}

Performing the matrix products of the controlled rotations gives
\begin{equation}
  U_A = \prod_{i,j=0}^{1} U_{i,j} = 
  \begin{pmatrix}
    c_{00} & 0 & 0 & 0 & -s_{00} & 0 & 0 & 0 \\
    0 & c_{01} & 0 & 0 & 0 & -s_{01} & 0 & 0 \\
    0 & 0 & c_{10} & 0 & 0 & 0 & -s_{10} & 0 \\
    0 & 0 & 0 & c_{11} & 0 & 0 & 0 & -s_{1,1}\\
    s_{00} & 0 & 0 & 0 &  c_{00} & 0 & 0 & 0 \\
    0 & s_{01} & 0 & 0 & 0 &  c_{01} & 0 & 0 \\
    0 & 0 & s_{10} & 0 & 0 & 0 & c_{10} & 0 \\
    0 & 0 & 0 & s_{11} & 0 & 0 & 0 &  c_{1,1}\\
  \end{pmatrix}
  =
  \begin{pmatrix}
      C & -S \\
      S &  C
  \end{pmatrix}
  \label{eqn-prod-Uij}
\end{equation}

Where $c_{ij} = \cos\frac{\theta_{ij}}{2}$ and 
$s_{ij} = \sin\frac{\theta_{ij}}{2}$.
Note that the controlled rotation operators commute and, hence,
the ordering of the products in \Cref{eqn-prod-Uij} is not important.
The final matrix in \Cref{eqn-prod-Uij} contains the 4x4 cosine and sine
blocks. 
The following matrices will also be expressed in terms of 4x4 blocks. 

The Hadamard operator on the row qubit is

\begin{equation}
  I \otimes H \otimes I = \frac{1}{\sqrt{2}}
  \begin{pmatrix}
    1 & 0 &  1 &  0 & 0 & 0 & 0 & 0\\
    0 & 1 &  0 &  1 & 0 & 0 & 0 & 0\\
    1 & 0 & -1 &  0 & 0 & 0 & 0 & 0\\
    0 & 1 &  0 & -1 & 0 & 0 & 0 & 0\\
    0 & 0 &  0 &  0 & 1 & 0 & 1 & 0\\
    0 & 0 &  0 &  0 & 0 & 1 & 0 & 1\\
    0 & 0 &  0 &  0 & 1 & 0 & -1 & 0\\
    0 & 0 &  0 &  0 & 0 & 1 & 0 & -1\\
  \end{pmatrix}
    =
  \begin{pmatrix}
      \Tilde{H} & 0 \\
      0  & \Tilde{H}
  \end{pmatrix}
  \label{eqn-IHI}
\end{equation}

The swap operator on the row and column qubits is:

\begin{equation}
  I \otimes SWAP =
  \begin{pmatrix}
    1 & 0 & 0 & 0 & 0 & 0 & 0 & 0\\
    0 & 0 & 1 & 0 & 0 & 0 & 0 & 0\\
    0 & 1 & 0 & 0 & 0 & 0 & 0 & 0\\
    0 & 0 & 0 & 1 & 0 & 0 & 0 & 0\\
    0 & 0 & 0 & 0 & 1 & 0 & 0 & 0\\
    0 & 0 & 0 & 0 & 0 & 0 & 1 & 0\\
    0 & 0 & 0 & 0 & 0 & 1 & 0 & 0\\
    0 & 0 & 0 & 0 & 0 & 0 & 0 & 1\\
  \end{pmatrix}
      =
  \begin{pmatrix}
      S_W & 0 \\
      0  & S_W
  \end{pmatrix}
  \label{eqn-ISWAP}
\end{equation}

Using the 4x4 block representations, accumulating the circuit operators gives:

\begin{equation}
  \begin{array}{rcl}
   \left( I \otimes H \otimes I \right)
   \left( I \otimes SWAP \right)
   U_A
   \left( I \otimes H \otimes I \right)  & = &
  \begin{pmatrix}
      \Tilde{H} & 0 \\
      0  & \Tilde{H}
  \end{pmatrix}
  \begin{pmatrix}
      S_W  & 0 \\
      0  & S_W 
  \end{pmatrix}
  \begin{pmatrix}
      C & -S \\
      S &  C
  \end{pmatrix}  
  \begin{pmatrix}
      \Tilde{H} & 0 \\
      0  & \Tilde{H}
  \end{pmatrix} \\[20pt]
  & = &
  \begin{pmatrix}
      \Tilde{H} S_W  C \Tilde{H} & -\Tilde{H} S_W  S \Tilde{H} \\
      \Tilde{H} S_W  S \Tilde{H} &  \Tilde{H} S_W  C \Tilde{H} 
  \end{pmatrix} 
  \end{array}
   \label{eqn-HSWUH}
\end{equation} 

Evaluating the upper left 4x4 block gives:

\begin{equation}
  \begin{array}{rcl}
    \Tilde{H} S_W  C \Tilde{H} & = & \frac{1}{2}
   \begin{pmatrix}
    1 & 0 &  1 &  0 \\
    0 & 1 &  0 &  1 \\
    1 & 0 & -1 &  0 \\
    0 & 1 &  0 & -1 \\
  \end{pmatrix}
  \begin{pmatrix}
    1 & 0 & 0 & 0 \\
    0 & 0 & 1 & 0 \\
    0 & 1 & 0 & 0 \\
    0 & 0 & 0 & 1 \\
  \end{pmatrix}
  \begin{pmatrix}
    c_{00} & 0 & 0 & 0 \\
    0 & c_{01} & 0 & 0 \\
    0 & 0 & c_{10} & 0 \\
    0 & 0 & 0 & c_{11} \\
  \end{pmatrix}
  \begin{pmatrix}
    1 & 0 &  1 &  0 \\
    0 & 1 &  0 &  1 \\
    1 & 0 & -1 &  0 \\
    0 & 1 &  0 & -1 \\
  \end{pmatrix} \\[20pt]
  
  & = & \frac{1}{2}
  
   \begin{pmatrix}
    1 & 0 &  1 &  0 \\
    0 & 1 &  0 &  1 \\
    1 & 0 & -1 &  0 \\
    0 & 1 &  0 & -1 \\
  \end{pmatrix}
  \begin{pmatrix}
    1 & 0 & 0 & 0 \\
    0 & 0 & 1 & 0 \\
    0 & 1 & 0 & 0 \\
    0 & 0 & 0 & 1 \\
  \end{pmatrix}
  \begin{pmatrix}
    c_{00} & 0      & c_{00}  & 0       \\
    0      & c_{01} & 0       & c_{01}  \\
    c_{10} & 0      & -c_{10} & 0       \\
    0      & c_{11} & 0       & -c_{11} \\
  \end{pmatrix} \\[20pt]

  & = & \frac{1}{2}

  \begin{pmatrix}
    1 & 0 &  1 &  0 \\
    0 & 1 &  0 &  1 \\
    1 & 0 & -1 &  0 \\
    0 & 1 &  0 & -1 \\
  \end{pmatrix}
  \begin{pmatrix}
    c_{00} & 0      & c_{00}  & 0       \\
    c_{10} & 0      & -c_{10} & 0       \\
    0      & c_{01} & 0       & c_{01}  \\
    0      & c_{11} & 0       & -c_{11} \\
  \end{pmatrix} \\[20pt]

  & = & \frac{1}{2}

  \begin{pmatrix}
    c_{00} &  c_{01} &  c_{00} &  c_{01} \\
    c_{10} &  c_{11} & -c_{10} & -c_{11} \\
    c_{00} & -c_{01} &  c_{00} & -c_{01} \\
    c_{10} & -c_{11} & -c_{10} &  c_{11} \\
  \end{pmatrix}
  \end{array}
  \label{eqn-HSCH-block}
\end{equation}

Setting $\theta_{ij} = 2\cos^{-1}(a_{ij})$, the upper left 2x2 block
contains the original matrix with a subnormalisation factor of 2.

%
\subsection{Arcsin based encoding}
\label{subsec-app-arcsin}

\begin{figure}[ht]
  \centering
  \includegraphics[width=0.80\textwidth]{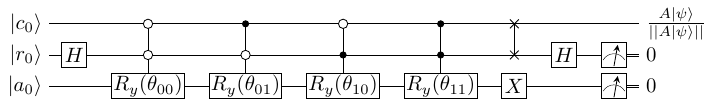}
  \caption{Circuit for full encoding of a 2x2 matrix using big-endian indexing with flipping of encoding block and 
  $\theta_{ij} = 2\sin^{-1}(2a_{ij})$.}
  \label{fig-app_2x2_orig_x_big}
\end{figure}

If instead of $I \otimes SWAP$ in \Cref{eqn-ISWAP}, $X \otimes SWAP$ is used,
see \Cref{fig-app_2x2_orig_x_big}, then \Cref{eqn-HSWUH} becomes:

\begin{equation}
  \begin{array}{rcl}
   \left( I \otimes H \otimes I \right)
   \left( X \otimes SWAP \right)
   U_A
   \left( I \otimes H \otimes I \right)  & = &
  \begin{pmatrix}
      \Tilde{H} & 0 \\
      0  & \Tilde{H}
  \end{pmatrix}
  \begin{pmatrix}
      0 & S_W  \\
     S_W  & 0
  \end{pmatrix}
  \begin{pmatrix}
      C & -S \\
      S &  C
  \end{pmatrix}  
  \begin{pmatrix}
      \Tilde{H} & 0 \\
      0  & \Tilde{H}
  \end{pmatrix} \\[20pt]
  & = &
  \begin{pmatrix}
    \Tilde{H} S_W  S \Tilde{H} &  \Tilde{H} S_W  C \Tilde{H} \\
    \Tilde{H} S_W  C \Tilde{H} & -\Tilde{H} S_W  S \Tilde{H}
  \end{pmatrix} 
  \end{array}
   \label{eqn-HXSWUH}
\end{equation}

Following the same steps as \Cref{eqn-HSCH-block}, the upper left block is now:

\begin{equation}
    \Tilde{H} S_W  S \Tilde{H}  =  \frac{1}{2}
      \begin{pmatrix}
    s_{00} &  s_{01} &  s_{00} &  s_{01} \\
    s_{10} &  s_{11} & -s_{10} & -s_{11} \\
    s_{00} & -s_{01} &  s_{00} & -s_{01} \\
    s_{10} & -s_{11} & -s_{10} &  s_{11} \\
  \end{pmatrix}
  \label{eqn-HSSH-block}
\end{equation}

Now, setting $\theta_{ij} = 2\sin^{-1}(a_{ij})$ gives the original matrix, 
again with a subnormalisation factor of 2. 
This is significant because sparse matrices with large number of
zeros will generate zero angle rotations that can be removed directly
without the \textsc{fable} approximations.
Indeed, the number of non-zero rotations is equal to the number
of non-zero entries in the original matrix.

%
\section{Emulating QSVT}
\label{app-sec-emulate}

Emulating QSVT on a classical computer can be somewhat
frustrating as exactly the same unitaries $U_A$ and 
$U_{A}^{\dagger}$ are applied repeatedly.
Computing and storing the unitaries is an effective strategy
for small numbers of qubits but the $\mathcal{O}(N^2)$ scaling of the
dense matrix-vector multiplication quickly obviates any advantage.
For relatively short encoding circuits, running the circuit in
emulation creates the situation where a sequence of 
sparse matrix-vector multiplications is faster than the 
corresponding single dense matrix-vector multiplication. 
In fact, this is always the case, but the cost of circuit emulation
soon becomes prohibitive too.
Whilst each application of the unitary or its adjoint may not be
prohibitive, it is repeating them 100s, 1,000s or even 10,000s
of times that becomes prohibitive. 
The highest degree polynomial used herein had 16,813 phase factors.

The following sub-sections discuss approaches for accelerating
QSVT emulators. Of course, QSVT will be run on the quantum computer
and the techniques discussed here have no bearing on the speed on
a quantum computer. The aim is to enable more rapid progress in
algorithm and application development.

All of the following improvements are implemented using big endian 
ordering. Implementation in little endian ordering is
straightforward.

\subsection{\textsc{prepare-select} encoding}
\label{app-subsec-emul-ps}

The first step to accelerate \textsc{prepare-select} encoding is to 
utilise the fact that the \textsc{select} operator,
\Cref{eqn-sel01}, produces a direct sum of the unitaries:

\begin{equation}
  S = 
  \begin{pmatrix}
     U_0 &     &        &         \\
         & U_1 &        &         \\
         &     & \ddots &         \\
         &     &        & U_{M-1} \\
  \end{pmatrix}
  \label{eqn-sel02}
\end{equation}

Applying $S$ to the qubits in QSVT circuit (assuming no other
qubits) has the block structure:

\begin{equation}
    I \otimes S = 
    \begin{pmatrix}
     S & 0  \\
     0 & S  \\
  \end{pmatrix}
  \label{eqn-sel03}
\end{equation}

Further, each unitary, $U_i$, is a tensor product of
1-sparse Pauli matrices. Hence, $S$ is also 1-sparse.
If $s_{ij}$ is the only entry on the $i^{th}$ row of $S$ then the
application of the unitary maps:
\begin{align}
  \label{eqn-sel-eff}
  \psi_i     &\mapsto s_{ij}\psi_j \\ \nonumber
  \psi_{i+N} &\mapsto s_{ij}\psi_{j+N} 
\end{align}

where $N$ is the number of rows in $S$.
This can be done as an efficient vector multiply using 
a temporary work vector.
Doing this using a sparse matrix data structure carries
significant overheads.
However, the column indices, which are assumed to be a variable
length list, but are actually just a vector, can be extracted
to directly perform the vector-vector multiplication.

If the encoding is used within a non-linear solver where
the sparsity pattern of the matrix does not change then
the Select operator can be constructed once at the outset
and reused. This cannot be done if unitaries with small
coefficients are trimmed as they is likely to remove
different unitaries on each non-linear iteration.

As the \textsc{prepare} operator only acts on the \textsc{prepare} register, 
it is not initially a bottleneck until the \textsc{select} operator
has been streamlined.

\begin{figure}[ht]
  \centering
  \includegraphics[width=0.90\textwidth]{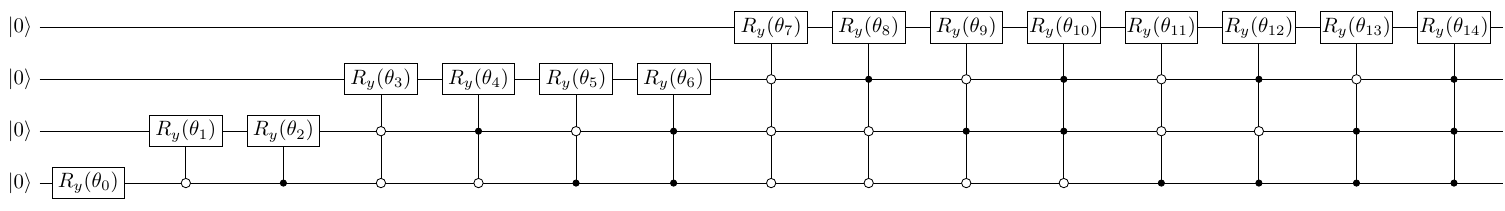}
  \caption{\textsc{prepare} circuit for loading 15 LCU coefficients, 
           big endian ordering.}
  \label{fig-araujo02}
\end{figure}

The \textsc{prepare} operator used here is a tree state loader
\cite{mottonen2004transformation}.
The levels in the tree correspond to levels in the circuit,
as shown in \Cref{fig-araujo02}.
Following \Cref{eqn-prod-Uij}, the operator for the
$2^{nd}$ level in the circuit is:

\begin{equation}
  U_2 = 
  \begin{pmatrix}
    c_3 & -s_3 & 0   & 0    & 0   & 0    & 0   & 0 \\
    s_3 &  c_3 & 0   & 0    & 0   & 0    & 0   & 0 \\
    0   & 0    & c_4 & -s_4 & 0   & 0    & 0   & 0 \\
    0   & 0    & s_4 & c_4  & 0   & 0    & 0   & 0 \\
    0   & 0    & 0   & 0    & c_5 & -s_5 & 0   & 0 \\
    0   & 0    & 0   & 0    & s_5 &  c_5 & 0   & 0 \\
    0   & 0    & 0   & 0    & 0   & 0    & c_6 & -s_6 \\
    0   & 0    & 0   & 0    & 0   & 0    & s_6 &  c_6\\
  \end{pmatrix}
  \label{eqn-prep-l2}
\end{equation}

Applying $U_2$ to full \textsc{prepare-select} register 
requires the a unitary of the form:

\begin{equation}
    I \otimes U_2 \otimes I^{\otimes m}
    \label{eqn-prep-l3}
\end{equation}
where $m$ is the number of qubits preceding the tree level in
the circuit.
This unitary is 2-sparse for all the levels in the tree
loader and has the same block structure as \Cref{eqn-sel03}
when applied to the whole circuit. 
This can be implemented using by extending the approach used
for the \textsc{select} operator:

\begin{align}
  \label{eqn-prep-eff}
  \psi_i     &\mapsto s_{ij}\psi_j     + s_{ik}\psi_k\\ \nonumber
  \psi_{i+N} &\mapsto s_{ij}\psi_{j+N} +  s_{ik}\psi_{k+N} 
\end{align}

This can be applied to each level of the state loader, but is
most beneficial for the lowest levels of the tree which have
large numbers of leaves.
In practice, levels with small numbers of rotations are
executed as a circuit, as this is more efficient than
accumulating the equivalent 2-sparse unitary using
\Cref{eqn-prep-l3}.

Implementing \Cref{eqn-prep-eff} and \Cref{eqn-sel-eff}
directly can lead to bespoke code that is harder to maintain.
If $U_2$ were the full \textsc{\textsc{prepare-select}} dense operator, then
expanding \Cref{eqn-prep-l3} to make a full circuit operator
can require large amounts of memory and be inefficient to
apply. A common alternative is to perform in-situ
matrix vector multiplication which strides through the
state vector and performs all the multiplications for 
each matrix entry. However, for 1-sparse and 2-sparse
matrices it is more efficient to create the full circuit
operator.

\subsection{Query Oracle encoding}
\label{app-subsec-emul-qo}

Efficiently implementing query oracle encoding follows
the same approach as separating the \textsc{prepare} and \textsc{select}
registers. From \Cref{app-2x2enc}, the query oracle 
has a 2x2 block diagonal form:

\begin{equation}
  O_A = 
  \begin{pmatrix}
    c_{0} & 0 & 0 & 0 & -s_{0} & 0 & 0 & 0 \\
    0 & c_{1} & 0 & 0 & 0 & -s_{1} & 0 & 0 \\
    0 & 0 & c_{2} & 0 & 0 & 0 & -s_{2} & 0 \\
    0 & 0 & 0 & c_{3} & 0 & 0 & 0 & -s_{3}\\
    s_{0} & 0 & 0 & 0 &  c_{0} & 0 & 0 & 0 \\
    0 & s_{1} & 0 & 0 & 0 &  c_{1} & 0 & 0 \\
    0 & 0 & s_{2} & 0 & 0 & 0 & c_{2} & 0 \\
    0 & 0 & 0 & s_{3} & 0 & 0 & 0 &  c_{3}\\
  \end{pmatrix}
  \label{eqn-oracle-Uij}
\end{equation}

When applied to the entire QSVT circuit it has the
the 4x4 block diagonal structure:

\begin{equation}
    I \otimes O_A = 
    \begin{pmatrix}
     C & -S & 0 & 0 \\
     S &  C & 0 & 0 \\
     0 &  0 & C & -S \\
     0 &  0 & S & C \\
  \end{pmatrix}
  \label{eqn-sel04}
\end{equation}

This enables an efficient in-situ implementation
using 4 local variables:

\begin{align}
    w_{i}      &= c_i\psi_i     - s_i\psi_{i+N/2} \\ \nonumber
    w_{i+N/2}  &= s_i\psi_i     + c_i\psi_{i+N/2} \\ \nonumber
    w_{i+N}    &= c_i\psi_{i+N} - s_i\psi_{i+3N/2} \\ \nonumber
    w_{i+3N/2} &= s_i\psi_{i+N} + c_i\psi_{i+3N/2} \\ \nonumber
    \psi_{i}      &= w_{i}       \\ \nonumber
    \psi_{i+N/2}  &= w_{i+N/2}   \\ \nonumber
    \psi_{i+N}    &= w_{i+N}     \\ \nonumber
    \psi_{i+3N/2} &= w_{i+3N/2}  \\ \nonumber
\end{align}

where $N$ is the number of rows in $O_A$.
An advantage of this formulation is that it is thread safe
and it can be easily parallelised using shared or distributed
memory programming.
The same is not true for the \textsc{select} operator,
\Cref{eqn-sel-eff}, as the Pauli strings do not, in general,
create a block diagonal structure.
For \textsc{arcsin} encoding the Pauli X gate s directly
implemented in $O_A$.

Each of the SWAP operations that follow the
query oracle can be efficiently created as a 1-sparse
permutation operator acting on the whole circuit.
These can be directly implemented as above.
Finally, the Hadamard gates are best implemented as individual
operations. Assembling tensor products of
Hadamard gates creates fully populated operators.
Each Hadamard gate is coded as a 2-sparse full circuit operator.
Streamlining the SWAP and Hadamard gates may seem like a small benefit,
but once the oracle has been optimised, these further reduced the run-time
by almost a factor of 4 on 9x9 coupled matrix.
Of course, making a fast emulator does not translate to
a fast implementation on a quantum device; nor, bring the runtime anywhere
near close to the original classical algorithm, but it does enable
algorithm research to progress.

\end{document}